\pdfoutput=1

\documentclass[11pt]{article}

\usepackage[final]{acl}
\usepackage{subfigure}
\usepackage{booktabs}
\usepackage{adjustbox}
\usepackage[most]{tcolorbox}
\usepackage{times}
\usepackage{amssymb}
\usepackage{latexsym}
\usepackage{longtable}
\usepackage{amsmath}
\usepackage{multirow}
\usepackage[T1]{fontenc}

\usepackage[utf8]{inputenc}

\usepackage{microtype}

\usepackage{algorithm}
\usepackage{algorithmic}
\usepackage{inconsolata}

\usepackage{graphicx}

\title{Dynamic Simulation Framework for Disinformation Dissemination and Correction With Social Bots}

\author{
 \textbf{Boyu Qiao\textsuperscript{1,2}},
 \textbf{Kun Li\textsuperscript{1}},
 \textbf{Wei Zhou\textsuperscript{1}},
 \textbf{Songlin Hu\textsuperscript{1,2}},
\\
 \textsuperscript{1}Institute of Information Engineering, Chinese Academy of Sciences, \\
 \textsuperscript{2}School of Cyber Security, University of Chinese Academy of Sciences
\\
\texttt{qiaoboyu, likun2, zhouwei, husonglin@iie.ac.cn}
\\
}

\begin{document}

\maketitle

\begin{abstract}

In the "human-bot symbiotic" information ecosystem, social bots play key roles in spreading and correcting disinformation. Understanding their influence is essential for risk control and better governance. However, current studies often rely on simplistic user and network modeling, overlook the dynamic behavior of bots, and lack quantitative evaluation of correction strategies. To fill these gaps, we propose \textbf{MADD}, a \textbf{M}ulti-\textbf{A}gent-based framework for \textbf{D}isinformation \textbf{D}issemination. MADD constructs a more realistic propagation network by integrating the Barabási–Albert Model for scale-free topology and the Stochastic Block Model for community structures, while designing node attributes based on real-world user data. Furthermore, MADD incorporates both malicious and legitimate bots, with their controlled dynamic participation allows for quantitative analysis of correction strategies. We evaluate MADD using individual and group-level metrics. We experimentally verify the real-world consistency of MADD's user attributes and network structure, and we simulate the dissemination of six disinformation topics, demonstrating the differential effects of fact-based and narrative-based correction strategies.

\end{abstract}

\section{Introduction}

In the era of human-bot symbiosis, social bots are prevalent on social networks and actively involved in information sharing \cite{alrhmoun2023emergent, cresci2020decade}. Malicious bots are strategically deployed to spread disinformation and disrupt healthy online discussions \cite{bhale2024malicious}. Conversely, legitimate bots are increasingly employed for the task of disinformation correction \cite{ferrara2023social, costello2024durably}. Therefore, developing robust network propagation simulation frameworks that quantitatively analyze the interaction between these opposing bots in information dissemination can help us better govern disinformation.

\begin{figure}[h]
  \centering
  \includegraphics[width=0.45\textwidth]{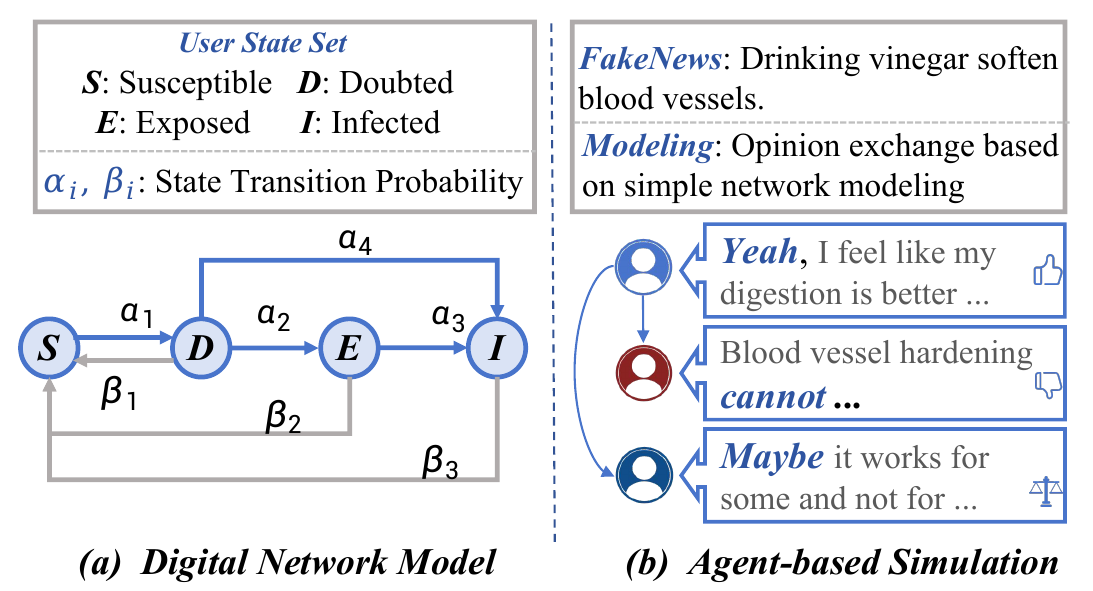}
  \centering
  \caption{(a) Digital network model: Simulates dynamic user state transitions using probability parameters $\alpha$ and $\beta$.
  (b) Agent-based simulation model: Dynamically update and disseminate the opinions of user agents in a simple dissemination network.}
  \label{fig:1}
  \end{figure}

  \vspace{-12pt}

Prior work in simulating disinformation propagation mainly follows two main paradigms: digital network-based propagation dynamics modeling and large language model (LLM)-based agent simulation. As shown in Figure \ref{fig:1}(a), digital network-based models \cite{pastor2015epidemic, gopal2022sepns, govindankutty2023fake} simulate information diffusion through user state transitions. Although computationally efficient, this conventional approach oversimplifies reality by ignoring user demographics and individual behavioral differences, limiting its ability to capture key phenomena such as repeated exposure and group polarization. Figure \ref{fig:1}(b) illustrates the emerging LLM-based agent simulation paradigm \cite{liu2024tiny, liu2024skepticism}, which improves user behavior modeling through semantic features and opinion propagating. However, it still faces notable limitations, redundant and irrelevant user attributes interfere with core propagation modeling, while inaccurate reconstruction of real-world network topologies causes deviations from actual scenarios.

Furthermore, current simulation research on social bots' involvement in disinformation dissemination and correction still faces significant limitations \cite{qiao2024botsim, averza2022evaluating}. For example, while Qiao et al. \shortcite{qiao2024botsim} simulate bot-driven disinformation spread, they neglect the construction of intervention scenarios and lack systematic effectiveness evaluation. These shortcomings highlight the urgent need for a more comprehensive simulation framework that integrates core demographic features, realistic propagating network topologies, and a robust evaluation metrics, thereby enhancing the authenticity of dissemination modeling.

To address the aforementioned limitations, we propose \textbf{MADD}, a \textbf{M}ulti-\textbf{A}gent-based framework for \textbf{D}isinformation \textbf{D}issemination, designed to model both the dissemination and correction of disinformation by social bots. MADD integrates the scale-free property of Barabási-Albert Model (BAM, \cite{schweimer2022generating}) and the community structure of Stochastic Block Model (SBM \cite{abbe2018community}) to construct a more realistic propagation network. It includes five node attributes, carefully designed based on real-world user data. Furthermore, we design three types of agents: regular users, malicious bots (spreading disinformation), and legitimate bots (correcting disinformation), enabling dynamic interactions. Using group and individual-level evaluation metrics, we validate the framework's effectiveness across six diverse disinformation topics and demonstrate the practical efficacy of fact-based and narrative-based correction strategies. Our study contributes the following:

  (1) We propose an innovative multi-agent framework for disinformation dissemination, featuring refined real user attributes and a realistic propagation network built by integrating BAM and SBM.

  (2) To the best of our knowledge, this is the first work to model the dynamic interplay between malicious and legitimate bots in a unified simulation. We innovatively evaluate and compare the effectiveness of two correction strategies. Furthermore, MADD's modular design enables its easy extension to test diverse correction strategies.

  (3) We meticulously design quantitative evaluation metrics at individual and group levels to systematically assess the impact of social bots on both the dissemination and correction of disinformation.


\section{Related Work}


\subsection{Disinformation Dissemination Modeling}

Disinformation dissemination modeling is theoretically significant for cybersecurity defense and information ecosystem security \cite{lopez2024frameworks}. Existing approaches include three main types: \textbf{digital network, agent-based simulation, and bot-assisted modeling.}

\textbf{Digital network modeling} captures \textbf{macro-level propagation characteristics} by defining node states and transition probabilities \cite{cifuentes2022mathematical}, using epidemic models like SIR (Susceptible, Infected, Recovered) \cite{zhu2017rumor}, SIS (Susceptible, Infected, Susceptible) \cite{dong2018sis}, SEIR (Susceptible, Exposed, Infected, Recovered) \cite{liu2018rumor}, and SEDIS (Susceptible, Exposed, Disseminated, Infected) \cite{govindankutty2022sedis}. However, these models \textbf{ignore memory effects and lack granular user attribute characterization, limiting the simulation of phenomena like repeated exposure.}

\textbf{Agent-based modeling (ABM)} addresses these limitations at the \textbf{micro level} by finely characterizing individual behavior and interactions, including user attributes and memory \cite{liu2024tiny, liu2024skepticism, muhammad2024agent}. By integrating role settings with dynamic memory-feedback, ABM enables detailed simulations of user cognition and social interactions. However, ABM still faces challenges: \textbf{irrelevant and redundant user attributes can reduce accuracy, and the lack of a systematic network model hinders capturing real-world topology and dynamics.}

\textbf{Bot-assisted dissemination modeling} further considers \textbf{automated accounts' impact on disinformation.} For example, Qiao et al. \shortcite{qiao2024botsim} simulate bot-driven spread but \textbf{lack intervention scenarios and impact assessment.} Aerza et al. \shortcite{averza2022evaluating} quantify user susceptibility using belief scores but typically \textbf{overlook the interplay of user attitudes and socio-environmental factors.}


\vspace{-5pt}

\subsection{Disinformation Correction Strategies}
Existing research primarily focuses on two correction strategies: \textbf{fact-based} and \textbf{narrative-based}. \textbf{Fact-based correction} \cite{boukes2023fighting, nyhan2020taking} widely used by professional fact-checkers, directly refutes disinformation with accurate facts and evidence through rational argumentation. \textbf{Narrative-based correction} \cite{dahlstrom2021narrative, vafeiadis2021fake} conveys truth by sharing authentic eyewitness accounts or related stories, enhancing immersion and persuasiveness. This study we systematically evaluate the effectiveness of legitimate bots applying these strategies across six disinformation topics.

\begin{figure*}[h]
  \centering
  \includegraphics[width=0.85\textwidth]{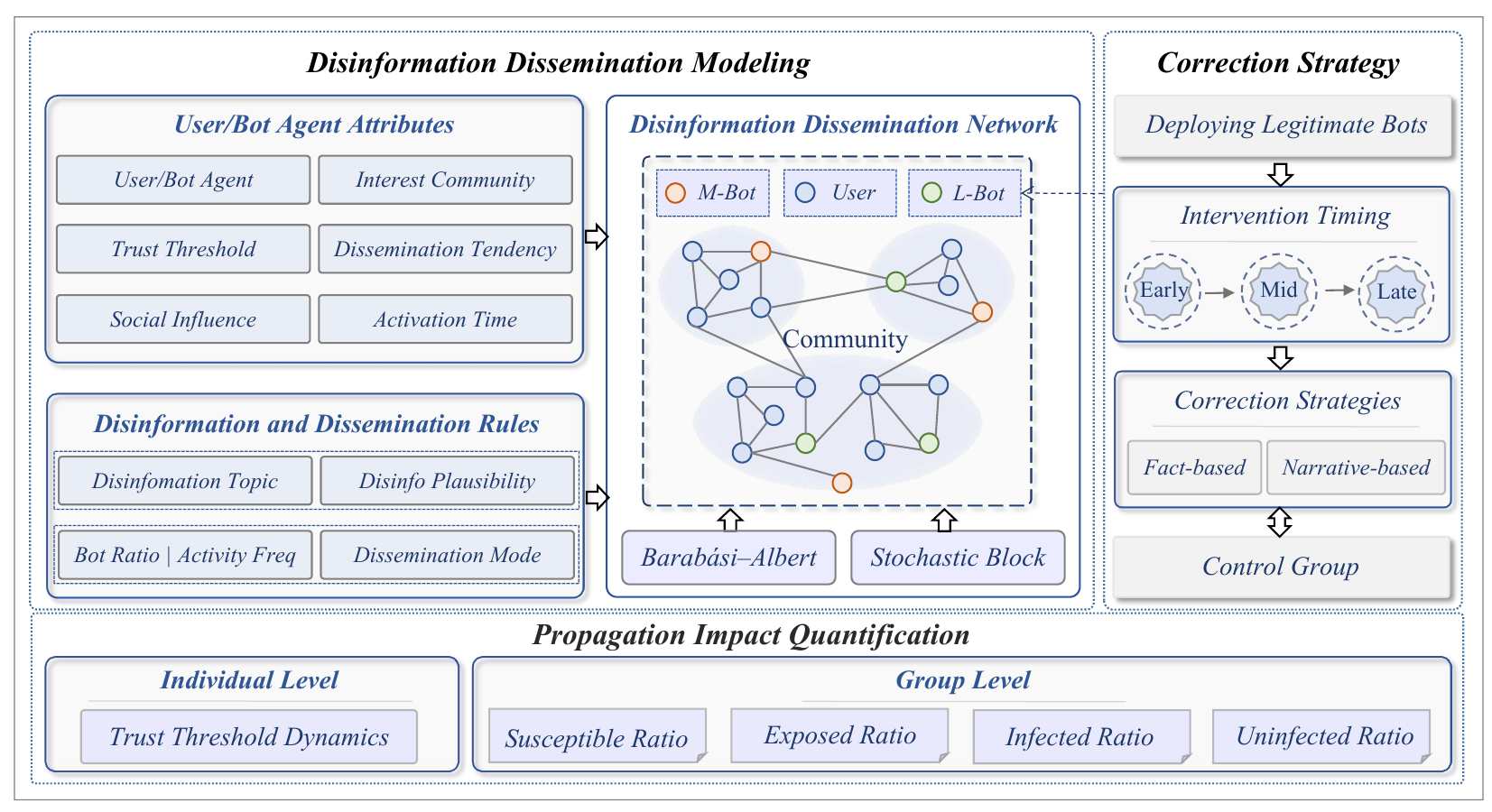}
  \centering
  \caption{Architecture of our proposed MADD. Appendix Figure \ref{fig:20} shows our extensions to the disinformation dissemination modeling.}
  \label{fig:2}
\end{figure*}


\section{Methodology}

In this work, we design a \textbf{M}ulti-\textbf{A}gent-based framework for \textbf{D}isinformation \textbf{D}issemination (MADD) to model the dynamic impact of social bots on disinformation spread and correction. As shown in Figure \ref{fig:2}, MADD comprises three modules: \textbf{disinformation dissemination modeling, correction strategy, and propagation impact quantification}.


\subsection{Disinformation Dissemination Modeling}

This module consists of three components: user/bot agent attributes, disinformation and  dissemination rules, and disinformation dissemination network.

\subsubsection{User/Bot Agent Attributes}

We design three agent types for the dissemination simulation: regular users, malicious bots (MBots), and legitimate bots (LBots). For efficiency and to reduce irrelevant attribute interference, we define five core attributes related to disinformation spread: interest community, trust threshold, dissemination tendency, social influence, and activation time.

\textbf{User/Bot Agents}: For regular users, we initialize their attributes based on real X/Twitter user data (data collection details in Appendix \ref{sec:B.1}). For MBots and LBots, their behavior is primarily controlled by automation, we configure their attributes using predefined procedures.

\textbf{Interest Community (IC):} To capture each user's communities of interest, we introduce interest community scores $\mathcal{IC}_{uj}$ for each user $u$ in community $j$. This score, ranging from 1 to 10, measures attention to community-specific content using LLM evaluation (complete prompts in Appendix \ref{sec:B.2}). It serves as a basis for community partitioning in subsequent propagation network construction and as a key feature for computing dissemination tendencies.

\textbf{Trust Threshold (TT):} To quantify user ability to identify disinformation in different communities, we use LLMs to evaluate the trust threshold $\mathcal{TT}_{uj}$ for each user $u$ within community $j$. By analyzing users' historical info and basic attributes via prompt engineering (complete prompt in Appendix \ref{sec:B.3}), we evaluate users’ resistance to disinformation in different communities.

\textbf{Dissemination Tendency (DT):} We combine the truncated power law distribution \cite{cha2010measuring, kwak2010twitter}) and user interest community scores $\mathcal{IC}_{uj}$ to quantify the user's dissemination tendency among communities. We fit the truncated power law distribution based on real user data, and we consider the user interest community score because user interest influences their likelihood to share information \cite{zhao2017impact}. The final dissemination tendency $\mathcal{DT}_{uj}$ for user $u$ in community $j$ is a weighted linear combination:

\vspace{-8pt}

\begin{equation}
  \label{equ:1}
  \begin{split}
    \mathcal{DT}_{uj} = & \left[ \theta \cdot \mathcal{CDF}\left(C \cdot (x_u)^{-\alpha} e^{-\lambda} \right) \right. \\
    & \left. + (1-\theta) \cdot \frac{\mathcal{IC}_{uj}}{ \max_{j} \mathcal{IC}_{uj}} \right] \cdot e^{- \xi n}, 
  \end{split}
\end{equation}


\noindent where $\theta \in [0,1]$ balances the truncated power-law component (capturing scale-free propagation patterns via its cumulative density function (CDF) with normalization constant $C$, share number $x_u > x_{min}$, exponent $\alpha$, and cutoff $\lambda$) and the max-normalized $\mathcal{IC}_{uj}$ (which preserves relative differences across communities). The exponential term $e^{-\xi n}$ models the decreasing tendency to share as a user encounters the same information more frequently. $(\alpha, \lambda, C)$ are optimized against real-world data (fitting details in Appendix \ref{sec:B.4}).

\textbf{Social Influence (SI):} To quantify the user's influence in the disinformation dissemination network, we introduce a social influence attribute to calculate the node degree centrality in the subsequently network. For a community $j$ with $n$ users, we collect each user $u$'s follower count $f_u$, forming the set $F = \{f_1, f_2, \dots, f_n\}$. Through normalization, the social influence $\mathcal{SI}_{uj}$ of user $u$ in community $j$ is defined as: $\mathcal{SI}_{uj} = \frac{f_u}{\sum_{i=1}^{n} f_i}$. A higher $\mathcal{SI}_{uj}$ indicates greater centrality and a stronger role in the dissemination process.

\textbf{Activation Time (AT):}  To capture the temporal dynamics of user participation in disinformation dissemination, we employ a discrete-time sequence design with $L$ steps, forming a time sequence $T=\{t_1, t_2, ..., t_L\}$. Based on statistical analysis of real-world user activity times, we define a time-varying activation probability $\mathcal{AT}_{ut} \in [0,1]$ for each user $u$, where a higher value indicates an increased likelihood of exposure to disinformation at that time step.

\subsubsection{Disinformation and Dissemination Rules}

We set disinformation by topic and plausibility. Dissemination rules is defined by bot ratio and activity frequency, and dissemination mode.

\textbf{Disinformation Topic:} The topic type of disinformation is associated with the community structure in the propagation network. Given that users in different communities show variations in their trust thresholds and tendency to spread specific disinformation topics, we introduce the attribute of disinformation topic.


\textbf{Disinformation Plausibility (DP):} DP refers to the degree of logical coherence and argumentative structure of disinformation, where higher plausibility increases the likelihood of users misinterpreting it as truthful. Inspired by Wan et al. \shortcite{wan2024dell}, we quantify plausibility $\mathcal{DP}_j$ of disinformation $j$ by evaluating emotional expression, propaganda strategies, information framing and logical consistency (prompt design in Appendix \ref{sec:B.6}).

\textbf{Bot Ratio and Activity Frequency}: To quantify the dynamic impact of social bots, we adjust malicious ($\mathcal{MR}_j$) and legitimate ($\mathcal{LR}_j$) bot injection ratios and their activity frequencies ($\mathcal{MF}_j$ and $\mathcal{LF}_j$) within each community $j$. This setup helps evaluate the potential impact of different bot manipulation strategies.

\textbf{Dissemination Mode}: To better model disinformation spread and minimize noise from redundant data, we focus on "reposts" and "quotes" as key indicators of user engagement. We treat direct reposts as implicit endorsements and analyze quoted content using LLMs to infer user stance. More detailed user interaction and decision-making strategies are introduced in Appendix B.11.

\vspace{-5pt}

\subsubsection{Disinformation Dissemination Network}
\vspace{-5pt}
We combine the Stochastic Block Model (SBM) \cite{abbe2018community,nowicki2001estimation} and Barabási-Albert Model (BAM) \cite{schweimer2022generating,barabasi1999emergence} to construct disinformation propagation networks. The SBM captures community structure with dense intra-community and sparse inter-community connections, while BAM introduces the scale-free properties and preferential attachment in real networks. Our construction process first assigns users to communities using interest community scores $\mathcal{IC}_{uj}$, then generate scale-free subgraphs within communities based on social influence scores $\mathcal{SI}_{uj}$. The resulting network naturally emerges with influential users (high-degree) as dissemination hubs and ordinary users (low-degree) as peripheral participants. (Implementation details in Appendix \ref{sec:B.7}.)

Our propagation network represents users as nodes and potential information diffusion pathways as edges. By adjusting parameters such as community assignment and the social influence of central nodes, we construct a network that simultaneously captures community structure and scale-free properties, providing a structured foundation for subsequent disinformation dissemination simulations.

\vspace{-5pt}
\subsection{Correction Strategy}
\vspace{-5pt}

We design \textbf{fact-based} and \textbf{narrative-based} correction strategies, to assess intervention effectiveness across communities. Specifically, we deploy malicious bots to spread disinformation and legitimate bots to deliver correction information. By comparing against \textbf{control groups}, we evaluate the effectiveness of different \textbf{correction strategies} and \textbf{intervention timings}.

\textbf{Correction Strategy: } To systematically evaluate effectiveness of different correction strategies, we manually design guiding content for two types of corrective messages, which can subsequently be rewritten and disseminated by legitimate bots with the assistance of LLMs (examples of both strategies are provided in Appendix \ref{sec:B.8}):

(1) \textbf{Fact-based Correction}: Directly refuting disinformation by citing authoritative data, such as verified data and scientific evidence.  

(2) \textbf{Narrative-based Correction}: Weakening the emotional appeal of disinformation through emotionally engaging narratives, such as personal stories and metaphors.

\textbf{Intervention Timing}: To examine the impact of intervention timing, we define three intervention stages: early, mid-stage, and late, corresponding to different time steps in the disinformation diffusion process (details in Appendix \ref{sec:B.9}).

\textbf{Control Group}: To evaluate the effects of the two correction strategies and three intervention timings, we establish a control group that does not implement any correction strategies. By comparing the propagation dynamics between the experimental and control groups, we quantify the effectiveness of different correction strategies across communities.

\subsection{Propagation Impact Quantification}
\vspace{-5pt}

We assess the impact of disinformation dissemination at both individual and group levels. The simulation and evaluation algorithm process is presented in Appendix \ref{sec:B.11}.

\vspace{-5pt}
\subsubsection{Individual Level}

\textbf{Trust Threshold Dynamics}: To evaluate the dynamic impact of repeated exposure to disinformation and corrective information on users' trust thresholds, we model changes in trust threshold using "enhancement" and "decay" terms. Specifically, we use the initial trust threshold $\mathcal{TT}_{uj}$ as a baseline and dynamically update the new trust threshold $\hat{\mathcal{TT}}_{uj}$ based on users' cumulative exposure to corrective and disinformation \cite{kemp2024role}. In modeling the dynamic adjustment of trust thresholds, we take into account both the user's individual interests and the influence of posts from neighboring users with varying levels of social influence. Considering the common psychological principle of diminishing marginal utility, we further describe the evolution of trust thresholds after multiple exposures as a process driven by both exponential growth and exponential decay \cite{daley1964epidemics, loomba2021measuring}:

\vspace{-8pt}


\begin{equation}
  \begin{aligned}
  & \mathrm{En}_{uj} = \gamma \left(1 - e^{ -\beta \sum_{k \in \mathcal{N}_{\mathrm{corr}}} \mathcal{SI}_k F'_{kj} } \right) \\
  & \mathrm{De}_{uj} = (1 - \gamma) \left(1 - e^{ -\delta \sum_{k \in \mathcal{N}_{\mathrm{dis}}} \mathcal{SI}_k F_{kj} } \right) \\
  & \hat{\mathcal{TT}}_{uj} = \mathrm{clip}\left( \mathcal{TT}_{uj} + \mathrm{En}_{uj} - \mathrm{De}_{uj}, 0, 1 \right),
  \end{aligned}
  \label{equ:2}
  \end{equation} 

 \vspace{-8pt}

\noindent where $\mathrm{En}_{uj}$ (Enhancement term) models the reinforcing effect from exposure to corrective content, while $\mathrm{De}_{uj}$ (Decay Term) captures the diminishing effect due to disinformation exposure. $\gamma$ balances correction and disinformation influence, while $\beta$ and $\delta$ control the changing rate. $\mathcal{N}_{\text{corr}}$ and $\mathcal{N}_{\text{dis}}$ are neighbors. $\mathcal{SI}_k$ is neighbor $k$'s social influence score. $F'_{kj}$ and $F_{kj}$ are the persuasive strength of corrective and disinformation content (evaluated by LLMs, Appendix \ref{sec:B.10}). Subsequently, user $u$'s disinformation discernment ability $\mathcal{DA}_{uj}$, which depends on their trust threshold $\hat{\mathcal{TT}}_{uj}$ and disinformation plausibility $\mathcal{DP}_j$, is calculated as follows:

\vspace{-8pt}

\begin{equation} 
  \label{equ:3} 
  \mathcal{DA}_{uj} = 1 - (1 - \hat{\mathcal{TT}}_{uj}) \mathcal{DP}_j.
\end{equation}

\subsubsection{Group Level}
At the group level, we assess the impact of disinformation using four user status ratios within community $j$ at time $t$: (1) \textbf{Susceptible Ratio} ($\mathcal{SR}_{t}^{j}$): Proportion of users unexposed to disinformation. (2) \textbf{Exposed Ratio} ($\mathcal{ER}_{t}^{j}$): Proportion of users having encountered disinformation at least once. (3) \textbf{Infected Spreaders Ratio} ($\mathcal{IR}_{t}^{j}$): Proportion of spreaders exposed to and believing disinformation, actively propagating it. (4) \textbf{Uninfected Spreaders Ratio} ($\mathcal{UR}_{t}^{j}$): Proportion of spreaders exposed to but not believing disinformation, distributing corrective information.

\vspace{-8pt}

\section{Experiments}

\vspace{-5pt}
\subsection{Experiment Setup}

We employ DeepSeek-V3 \cite{liu2024deepseek} as the base model, balancing response quality and computational efficiency. The experiment simulates six communities: \textbf{Business, Education, Entertainment, Politics, Sports, and Technology}. Regular user attributes originate from real user data, while bot agents simulate regular user behaviors, such as programmable activation time and dissemination tendency. Furthermore, we set up $15\%$ malicious bots \cite{ferrara2016rise} and $5\%$ legitimate bots in each community network to spread false and corrective information, respectively. Detailed parameter definitions, value ranges, and data sources are available in Appendix \ref{sec:A.1}.

\vspace{-5pt}
\subsection{MADD and Real-World Consistency Evaluation}

To confirm the MADD framework's trustworthiness, we perform a double verification: we first verify that user attributes and network configurations statistically match real social networks, and  and then we validate our simulations by comparing the spread trends of disinformation with existing empirical studies.

\begin{figure}[ht]
  \centering
  \subfigure[Interest Community.]{
    \begin{minipage}[t]{0.22\textwidth}
        \centering
        \hspace{-0.2cm}
        \includegraphics[width=0.98\textwidth]{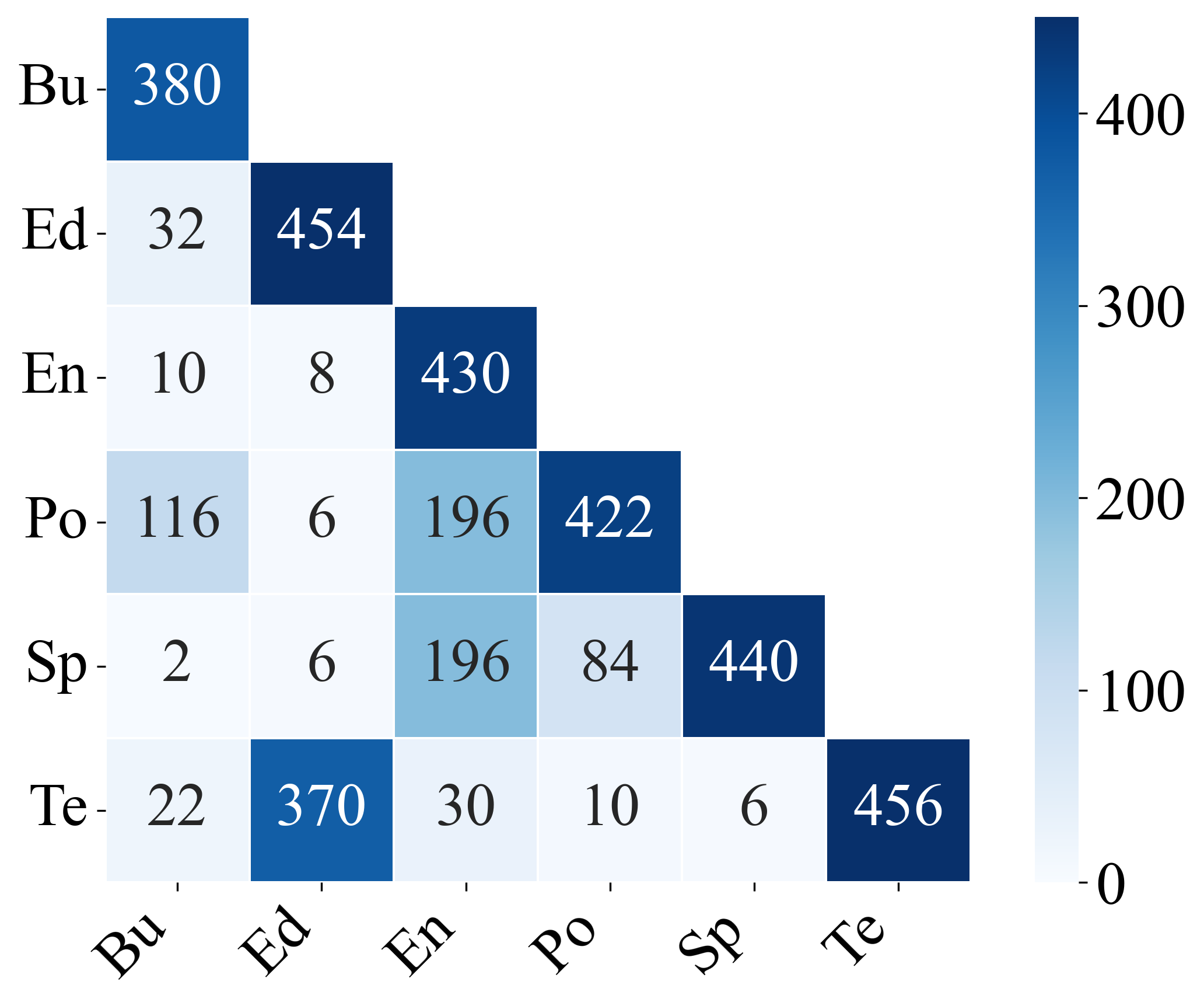}
    \end{minipage}
  }%
  \subfigure[Node Degree.]{
    \begin{minipage}[t]{0.22\textwidth}
        \centering
        \hspace{-0.2cm}
        \includegraphics[width=0.98\textwidth]{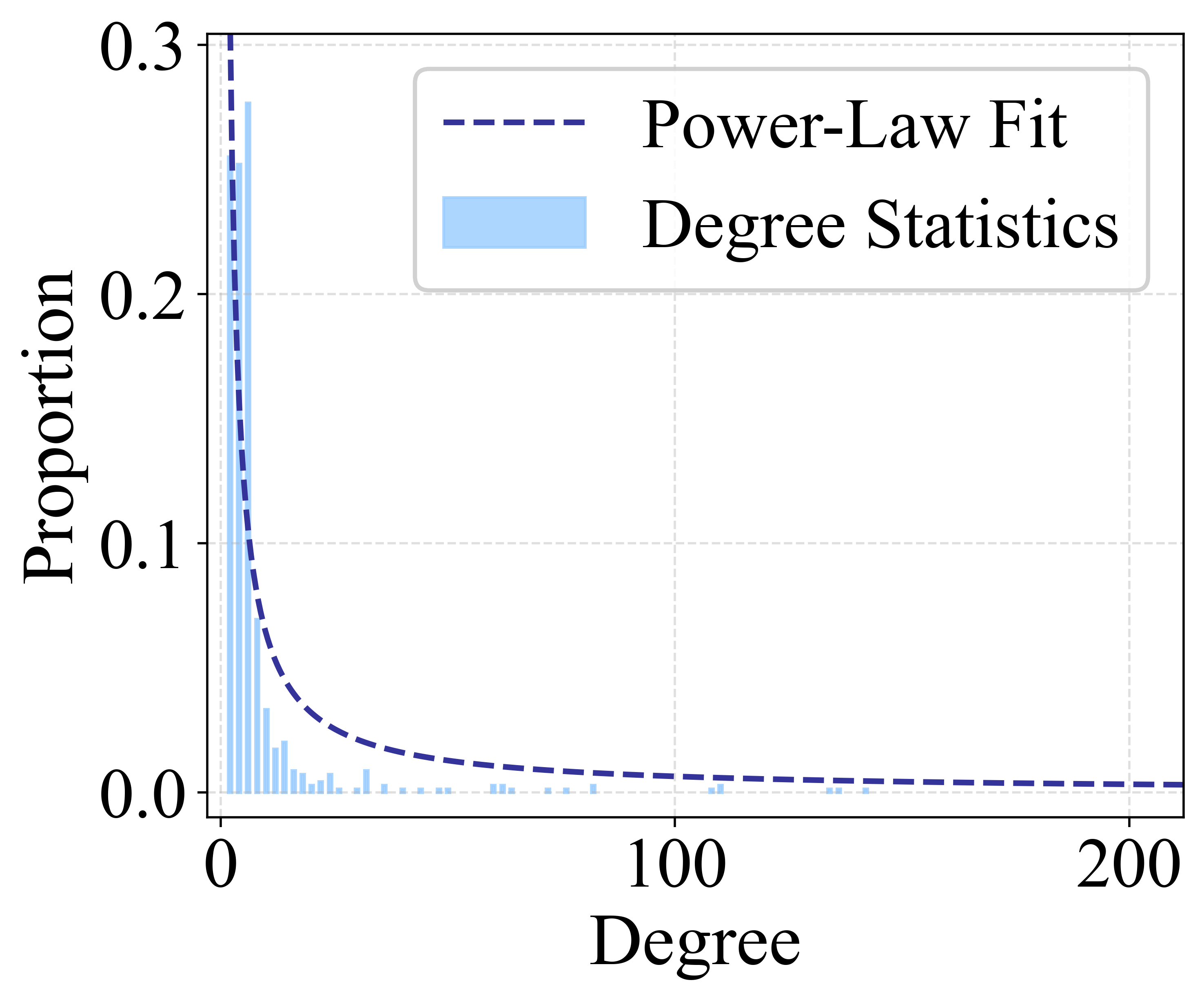}
    \end{minipage}
  }%
  \caption{Interest Community and Node Degree Distributions. In (a), the X and Y axis labels represent the first two letters of each of the six communities.}
  \label{fig:3}
\end{figure}

\textbf{Consistency of Attributes and Network Setup}: Figure \ref{fig:3}(a) shows that the distribution of user interest communities aligns with the structural characteristics of real-world social networks, exhibiting dense intra-community links and sparse inter-community links \cite{yang2012community}. Figure \ref{fig:3}(b) reveals that the network's degree distribution, based on real user social influence, follows a power law pattern. Furthermore, Appendix \ref{sec:C.1}, Figure \ref{fig:4} indicates users' dissemination tendencies conform to the heavy-tailed distribution observed in real networks \cite{goel2016structural}. Figure \ref{fig:5} shows user trust threshold scores approximately follow a normal distribution, consistent with the conclusion from previous studies that most users hold a neutral stance \cite{ecker2022psychological}.

\begin{figure}[ht]
  \centering
  \includegraphics[width=0.49\textwidth]{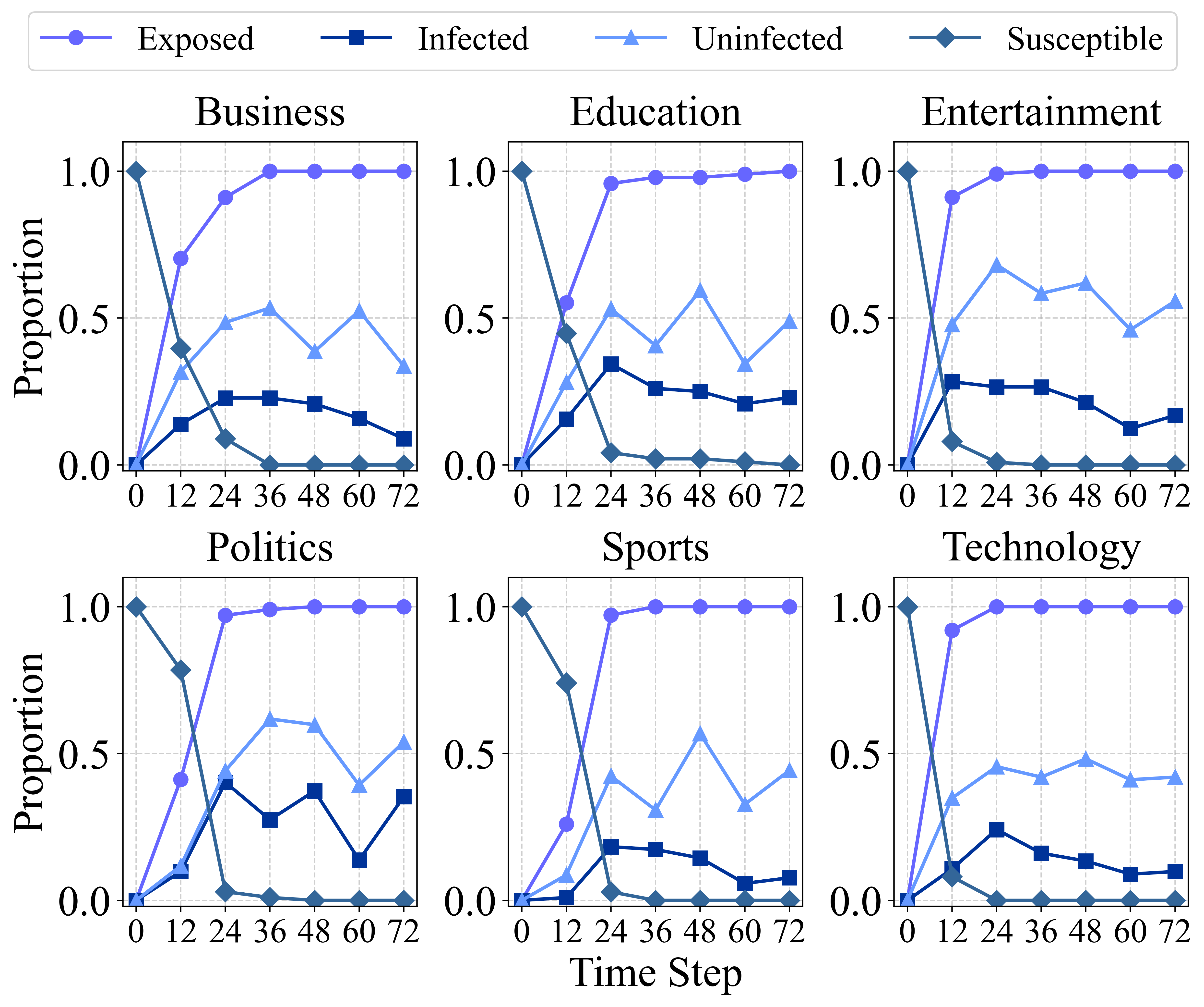}
  \caption{Proportion of user status in each community over time.}
  \label{fig:7}
\end{figure}

\begin{figure}[ht]
  \centering
  \includegraphics[width=0.49\textwidth]{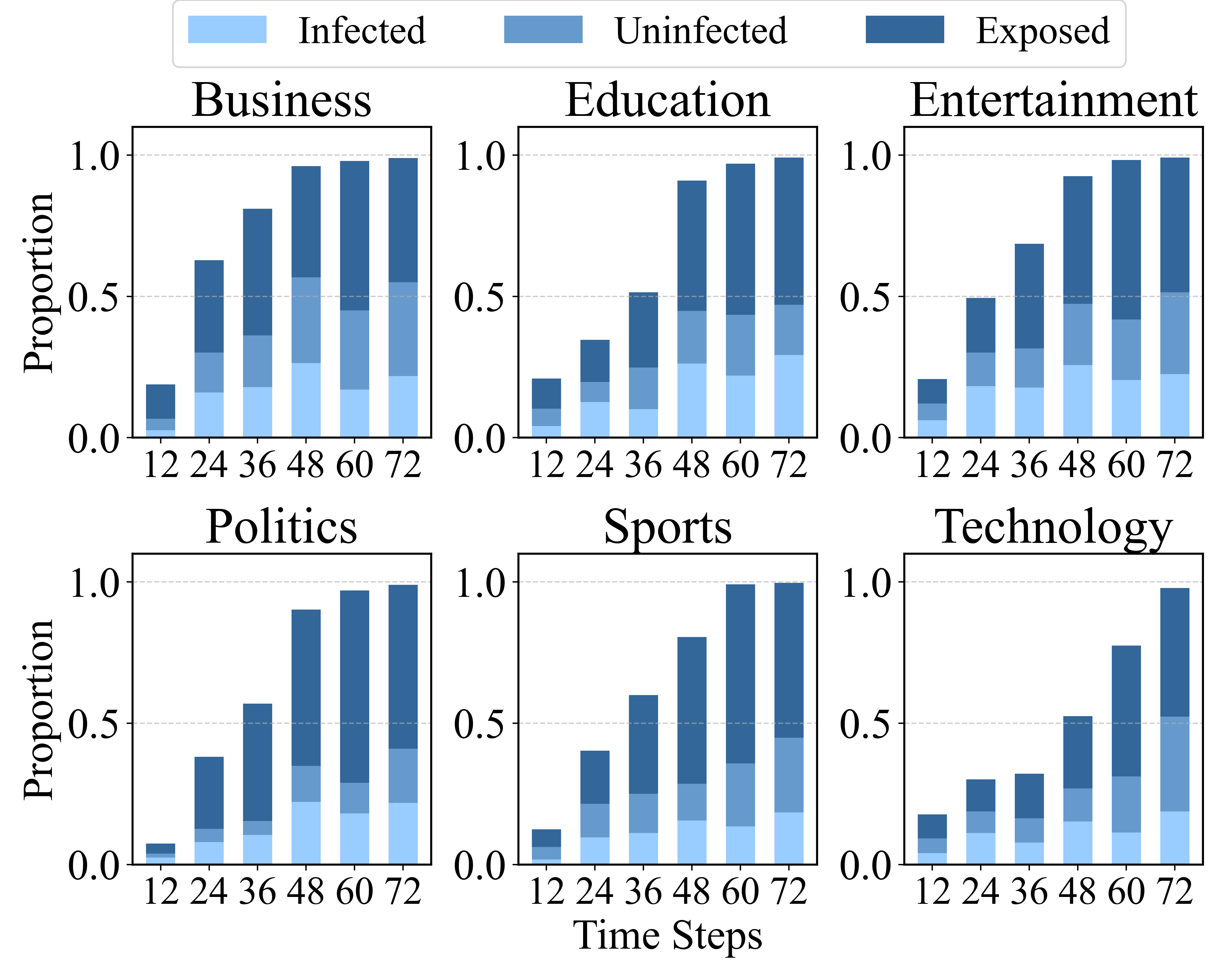}
  \caption{Proportion of user status change across the entire network over time.}
  \label{fig:8}
\end{figure}

\textbf{Simulation Results of MADD}: To validate the consistency of MADD's simulation with the real world, we simulate the diffusion of disinformation across six topics within the entire network constructed by six communities over 72 time steps. Each topic initiates within its most relevant community. Figure \ref{fig:7} illustrates the changing proportions of user statuses within each community (recorded every 12 time steps), while Figure \ref{fig:8} shows the disinformation's spread across the entire network. Notably, our model simulates user activation timing, leading to lower activity at time steps 12, 36, and 48 compared to 24, 48, and 72, which explains some of the fluctuations observed in Figure \ref{fig:7}.

Figure \ref{fig:7} shows that the proportion of "infected" spreaders in each community exhibits a trend of rapid initial increase followed by a gradual decline, peaking at approximately the 24-th time step. This aligns with existing empirical studies \cite{vosoughi2018spread,shao2018spread,tucker2018social} and classical epidemic spreading models like SIR \cite{govindankutty2024epidemic,gopal2022sepns}. Notably, the proportion of "uninfected" spreaders is higher than that of "infected" spreaders, indicating that a larger proportion of users can more successfully identify disinformation based on their trust thresholds and judgments about the plausibility of the disinformation. Furthermore, the peak time for the proportion of "uninfected" spreaders mostly occurs later than that of "infected" spreaders, suggesting that the spread of corrective information or rational responses is slower. In contrast, the "Politics" differs from other communities. The proportion of "infected" spreaders in this community does not exhibit the typical unimodal pattern but shows significant fluctuations at different time steps, likely due to the sensitive and debated nature of political information \cite{shin2018diffusion}. This simulation highlights the challenge of predicting political topics on social networks.

Figure \ref{fig:8} illustrates the evolving proportions of infected, uninfected, and exposed users across the entire network for different disinformation topics. Most topics take 60-72 time steps for network-wide spread. In contrast, within closely related communities (Figure \ref{fig:7}), spread is much faster, around 24 time steps. This suggests faster diffusion within communities but slower propagation across them due to the inherent community structure of the network. The increasing of all user states further suggests the ongoing diffusion of disinformation throughout the network. Capturing the critical turning point at which the "infected" rate begins to decline necessitates longer simulations. However, considering the computational cost of extended simulations, especially the LLM token usage for modeling user opinions in multi-level networks, subsequent experiments will focus on simulating spread within single communities to efficiently evaluate correction strategies.

\vspace{-10pt}
\subsection{Group-Level: Effectiveness Evaluation of Correction Strategies}

\begin{figure}[htb]
  \centering
  \includegraphics[width=0.49\textwidth]{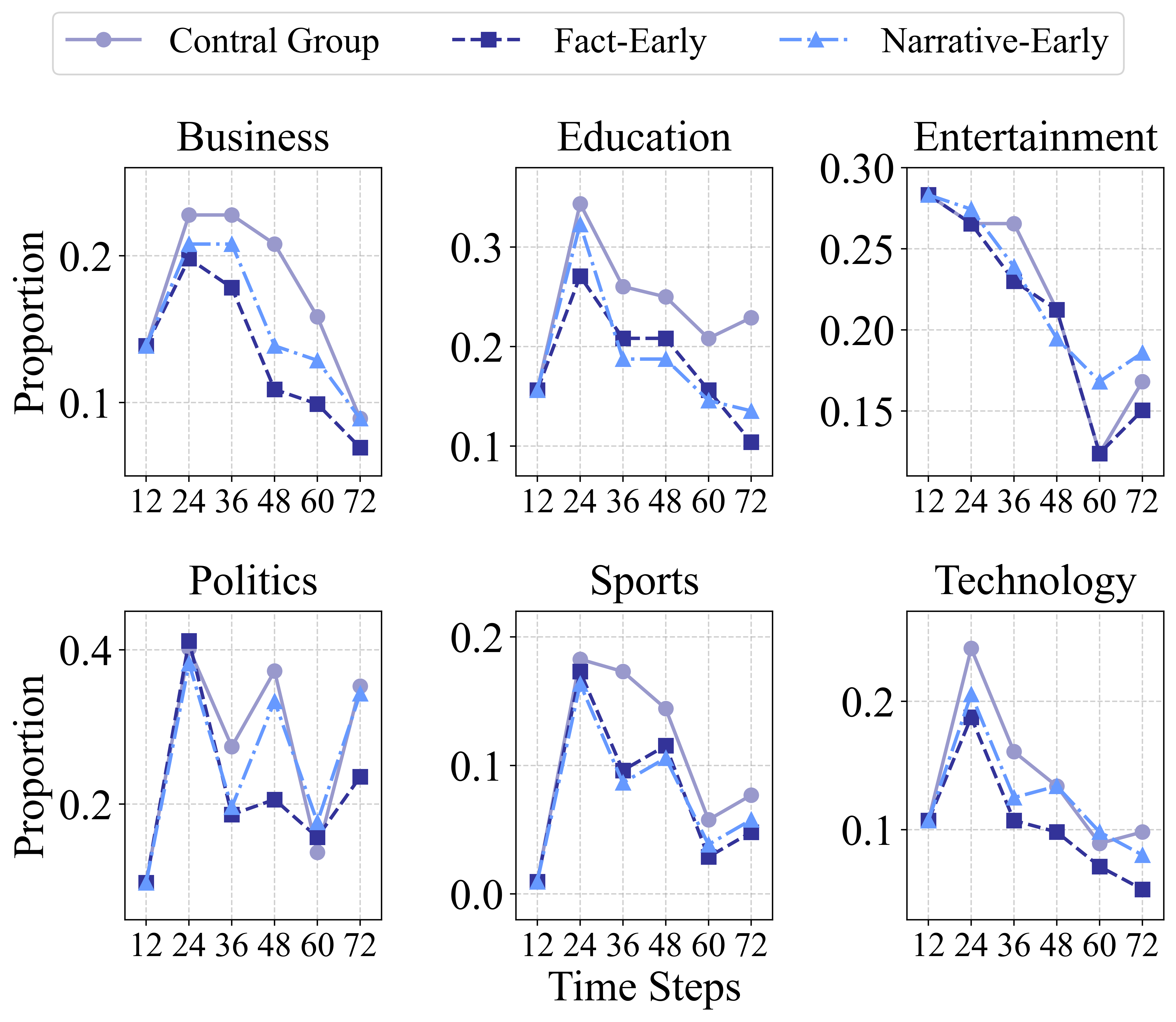}
  \caption{Comparison of early intervention effects of two correction strategies.}
  \label{fig:9}
\end{figure}

To evaluate two correction strategies, we implement them for early propagation in Figure \ref{fig:9} and compare the resulting changes in the proportion of infected spreaders over time against a control group (without external correction strategies). Observing Figure \ref{fig:9}, we find that fact-based correction is significantly more effective than narrative-based correction in `Business', `Politics', and `Technology'. For `Education' and `Sports', both strategies show comparable effectiveness in curbing disinformation spread. However, neither strategy significantly impacts `Entertainment', likely due to its inherent subjectivity and diverse interpretations, limiting the impact of debunking.

Appendix \ref{sec:C.2}, Figures \ref{fig:10} and \ref{fig:11} illustrate limited corrective effects of the two strategies in the mid and late propagation stages across most communities. Moreover, in certain communities like `Politics', we even observe negative impacts, likely due to substantial prior disinformation exposure causing users to distrust debunking and strengthen their original views, leading to an echo chamber \cite{garimella2018political}.

\subsection{Individual-Level: Dynamic Changes in User Trust Thresholds}

\begin{figure}[htb]
  \centering
  \includegraphics[width=0.49\textwidth]{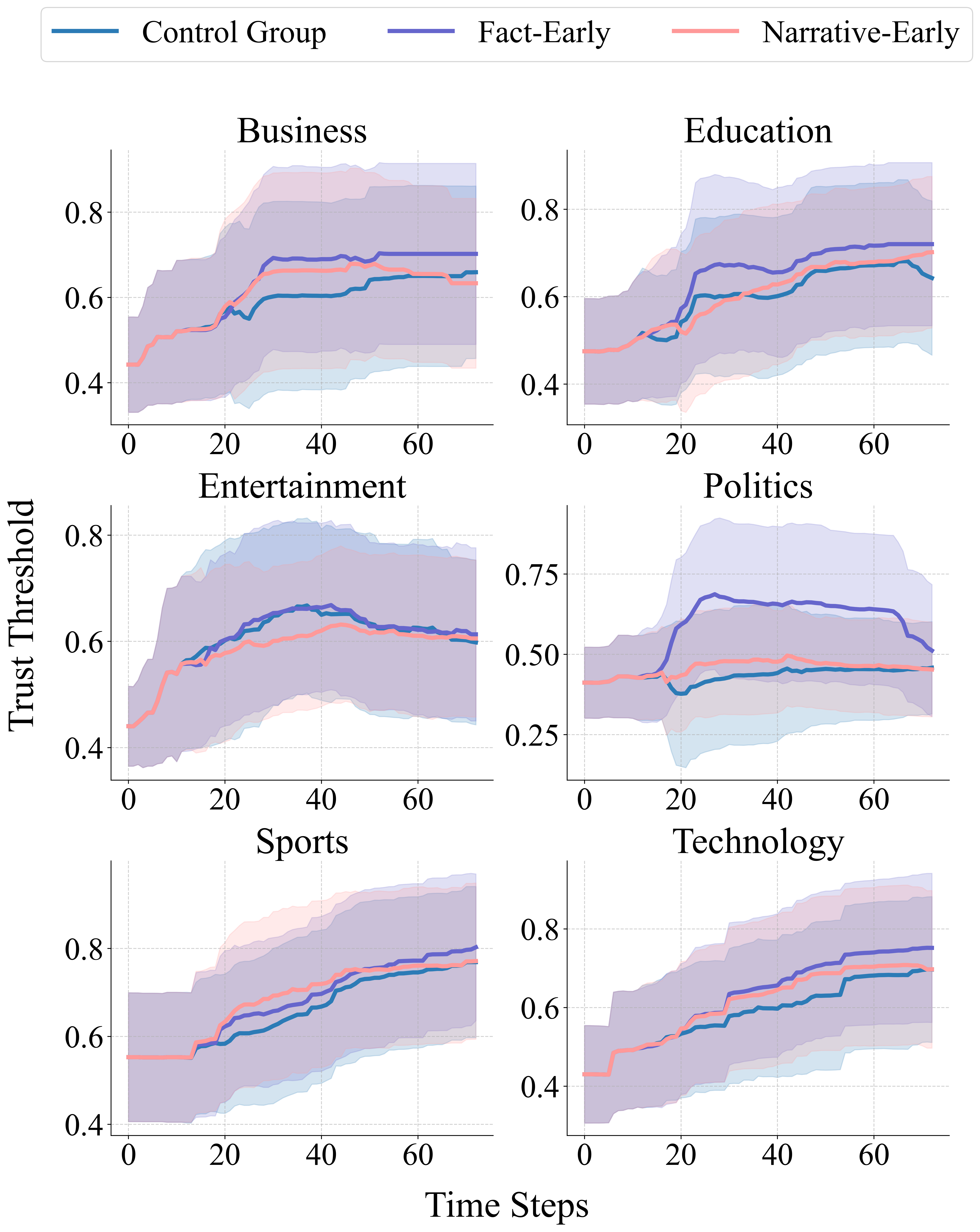}
  \caption{Effects of correction strategies on user trust thresholds during early-stage intervention}
  \label{fig:12}
\end{figure}

To assess the potential impact of different correction strategies on user trust thresholds, Figure \ref{fig:12} visualizes the dynamic evolution of user trust thresholds over time steps in the early intervention scenario, specifically displaying the mean and standard deviation of all user trust thresholds at each time step. Observing Figure \ref{fig:12}, we find that even in the control group where no external correction strategies are implemented, the user trust threshold for disinformation still exhibits a continuous upward trend, although its rate of increase is slower than in the simulations where correction strategies are applied. This observation indicates that even in the absence of intervention, a certain proportion of spontaneous debunking users exists within the network, who can mitigate the negative impact of disinformation to some extent.

Our further analysis comparing the impact of fact-based and narrative-based correction strategies on trust thresholds (relative to the control group) indicates that fact-based correction results in larger trust threshold increases across most communities. However, the "Politics" community displays a distinctive pattern, showing the smallest overall increase and the most significant fluctuations. Specifically, while fact-based corrective strategies significantly increased the trust threshold of this community initially, they eventually led to a downward trend in trust thresholds as disinformation continued to hedge against the spread of corrective information \cite{vosoughi2018spread}. This phenomenon may stem from the highly controversial nature of political information itself, coupled with the often greater emotional appeal of disinformation, prompting users to shift their positions repeatedly. In addition, the persistence of cognitive bias and group polarization in political communities \cite{van2021political} pose a substantial barrier to the effective dissemination and reception of corrective information.

Appendix \ref{sec:C.3}, Figures \ref{fig:13} and \ref{fig:14} illustrate the changes in user trust thresholds when correction strategies are implemented in the mid and late stages. We observe that their effect on increasing user trust thresholds for evaluations is extremely limited. Furthermore, in the "Entertainment," "Business," and "Politics" during the mid-stage, and in the "Sports" during the late-stage, the control group (without correction strategies) actually shows higher user trust thresholds than the experimental group (with correction strategies). This phenomenon likely stems from two main reasons: the intervention timing is delayed and echo chamber effects may have already formed within some user groups. Consequently, the aforementioned experimental results underscore the critical importance of intervening in the early stages of disinformation spread and implementing effective correction strategies.

\begin{figure}[h]
  \centering
  \subfigure[Trust Thresholds.]{
    \begin{minipage}[t]{0.22\textwidth}
        \centering
        \hspace{-0.2cm}
        \includegraphics[width=0.98\textwidth]{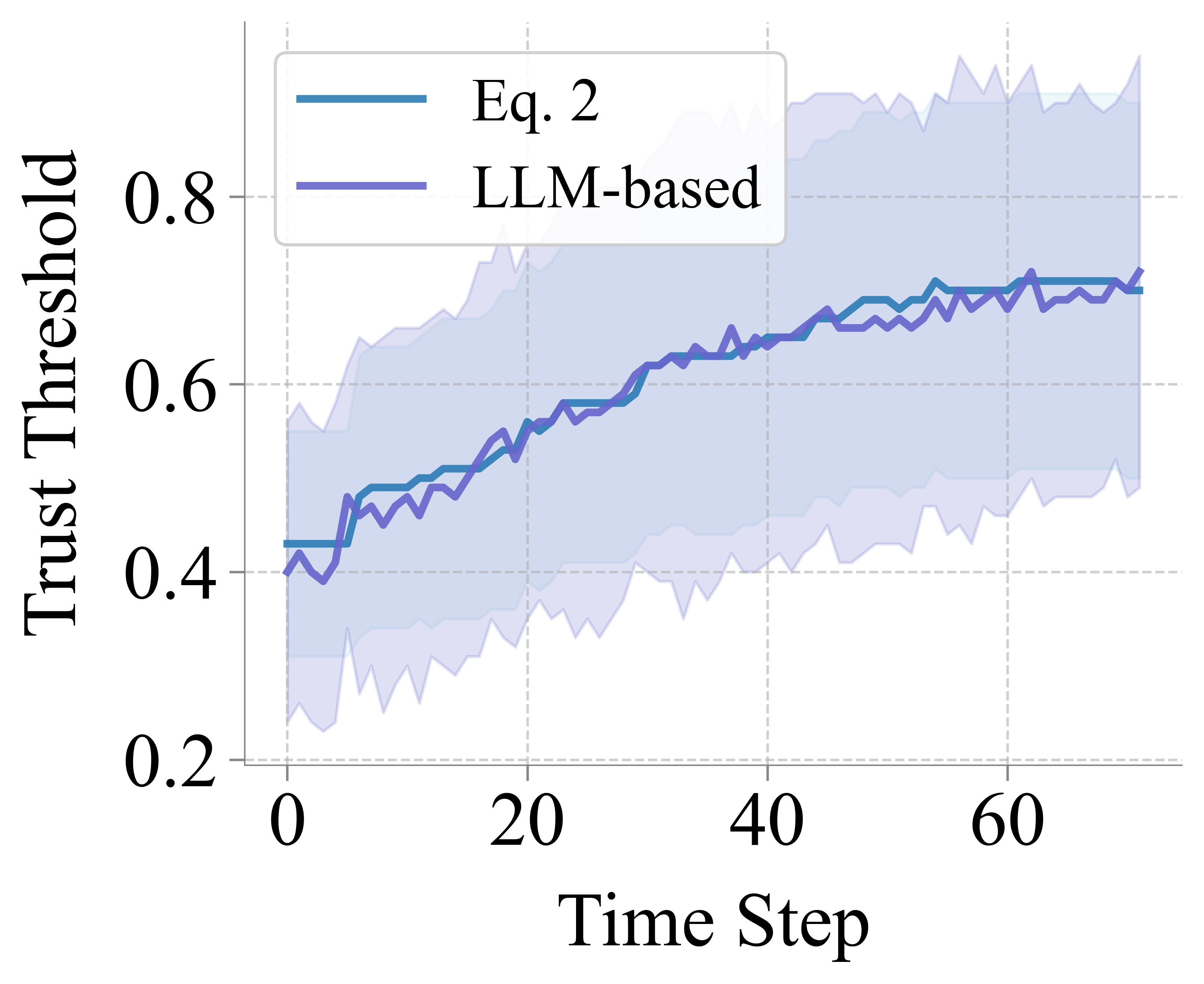}
    \end{minipage}
  }%
  \subfigure[Token Consumption.]{
    \begin{minipage}[t]{0.22\textwidth}
        \centering
        \hspace{-0.2cm}
        \includegraphics[width=0.98\textwidth]{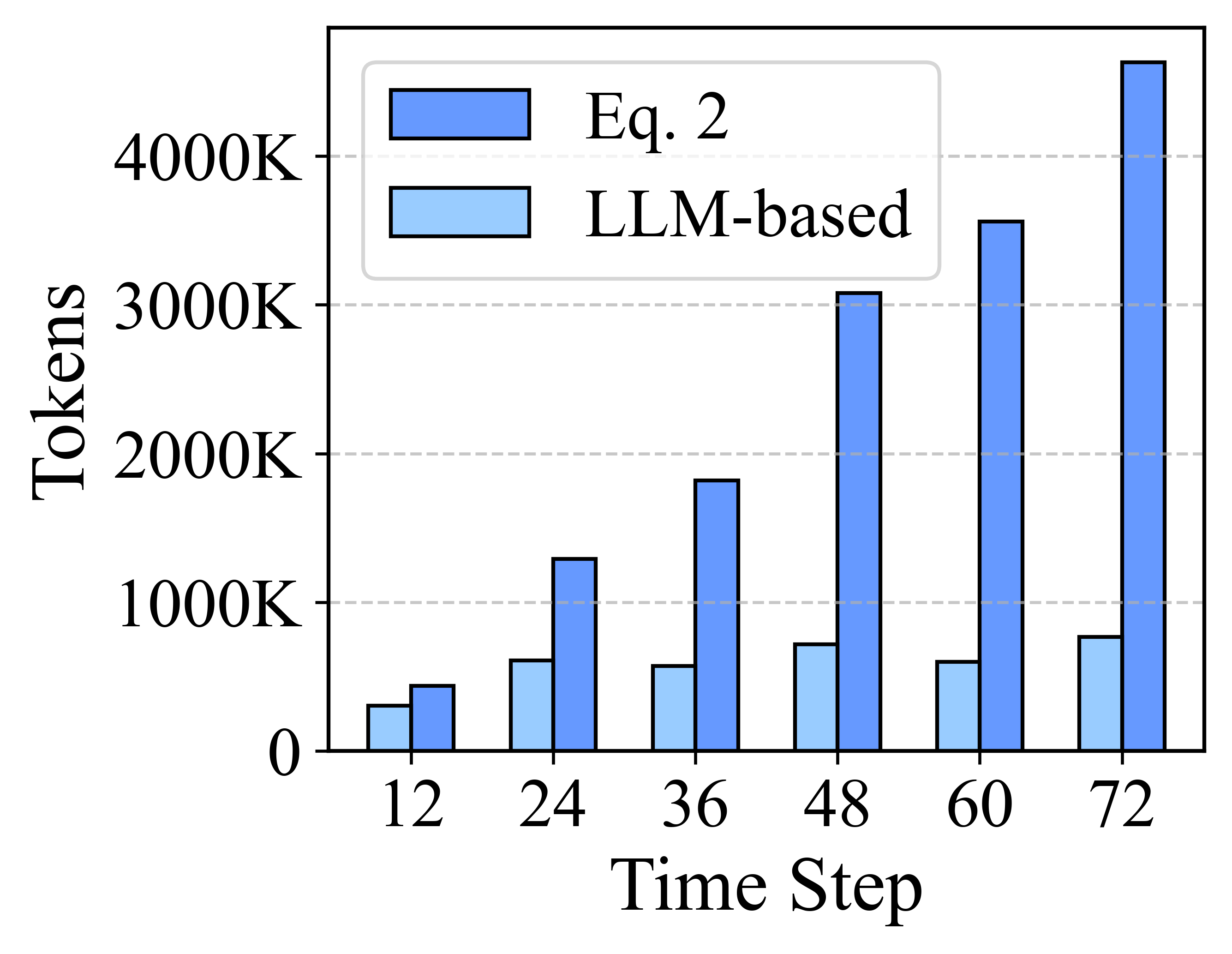}
    \end{minipage}
  }%
  \caption{Comparative Analysis of Trust Threshold Distributions and Token Consumption}
  \label{fig:15}
\end{figure}

\textbf{Trust Threshold Validity Assessment:} We validate the effectiveness of our proposed trust threshold calculation method (Equation \ref{equ:2}) by comparing it with an LLM-based evaluation method. Figure \ref{fig:15}(a) demonstrates a high degree of consistency in the trust threshold evolution trends of the both methods across different time steps. Furthermore, Figure \ref{fig:15}(b) reveals that our approach achieves a 76\% average reduction in token consumption compared to the LLM method. The experimental results indicate that MADD not only maintains evaluation accuracy comparable to LLMs but also significantly improves computational efficiency and effectively reduces resource overhead. These findings demonstrate the practical value of our proposed dynamic trust threshold estimation method in accurately characterizing user trust properties.

\section{Conclusion}

In this work, we address key limitations in current disinformation dissemination research, including oversimplified user and network models and the lack of quantitative evaluation of correction strategies. To overcome these challenges, we propose MADD. MADD simulates realistic online social dynamics by incorporating five key user attributes relevant to disinformation spread and modeling network structure with scale-free and community properties. We then utilize individual and group level metrics to rigorously assess how MBots spreading disinformation and LBots providing corrections impact regular users. Our comparative experiments demonstrate the effectiveness of various correction strategies in curbing disinformation, offering valuable insights for future research.

\section*{Limitations}
Our study has several limitations in the experimental design that need to be addressed. First, the experiment does not systematically test the differential effects of malicious bots and legitimate bots on ordinary user groups under varying ratio conditions. Second, due to constraints such as computational resources and token costs, we only simulate the evolution of user states over 72 time steps in the constructed propagation network, without conducting long-term dynamic simulations. Third, in evaluating the effectiveness of correction strategies, considering that a global network assessment would incur high token costs, we choose to conduct experimental validation within a single community network. Although this simplified approach meets the preliminary research needs, future studies could construct larger-scale network models and extend the simulation period. This would help more comprehensively assess the complex mechanisms of social bots in disinformation dissemination and correction processes, thereby providing a more reliable theoretical foundation for practical applications. Fourth, there remains a potential concern that the LLM used as a user agent may exhibit topic-specific biases, which could influence simulation outcomes. For example, even under identical simulation settings, the extent of information spread may vary across topics due to the LLM’s inherent preferences or assumptions. To mitigate such model-driven bias, future research should prioritize the selection of more neutral or balanced topics, thereby improving the objectivity and robustness of the simulation results.

\section*{Ethics Statement}

This study adheres to research ethics standards. First, the six topics of disinformation and their corresponding correction strategies used in this research are all selected from verified cases publicly released by internationally authoritative fact-checking agencies such as "Snopes," "Politifact," and "FactCheck." Our study does not generate any new disinformation content, ensuring full compliance with academic ethics requirements throughout the research process. Second, concerning data protection, for the real user data involved in the study, we implement rigorous anonymization measures to ensure the full protection of user privacy.

\bibliography{custom}

\clearpage

\appendix

\begin{table*}[ht]
  \centering
  \caption{Statistical information in each community. We present the mean and standard deviation of users' original posts, retweets, and quotes.} 
  \renewcommand\arraystretch{1.2}
  \adjustbox{width=0.95\textwidth}{%
  \begin{tabular}{ccccccc}
    \toprule[1pt]
  \textbf{Community}                  & \textbf{Entertainment} & \textbf{Technology} & \textbf{Sports} & \textbf{Business} & \textbf{Politics} & \textbf{Education} \\ \hline
  \textbf{Number of Users}            & 113                    & 112                 & 104             & 101               & 102               & 96                \\ 
  \textbf{Number of Posts}    & 24.05$\pm$21.54            & 14.38$\pm$13.09         & 31.70$\pm$24.25     & 34.17$\pm$21.17      & 36.01$\pm$28.96       & 23.96$\pm$15.65      \\ 
  \textbf{Number of Retweets} & 32.63$\pm$21.54            & 11.35$\pm$4.31         & 24.72$\pm$12.95     & 36.64$\pm$14.59      & 48.23$\pm$31.65       & 14.03$\pm$6.25      \\ 
  \textbf{Number of Quotes}   & 10.95$\pm$7.87            & 6.08$\pm$3.30         & 11.42$\pm$7.53     & 4.82$\pm$2.62       & 23.51$\pm$12.85       & 8.03$\pm$2.15       \\ \toprule[1pt]
  \end{tabular} }
  \label{table:1}
\end{table*}


\section{Experimental Parameter Settings}
\label{sec:A.1}

We present the parameter variables, their value ranges, definitions, and sources necessary for replicating the current simulation in Table \ref{table:2}, facilitating reproducibility for researchers. We construct six distinct community networks— \textbf{Entertainment, Technology, Sports, Business, Politics, and Education}, which collectively form the integrated propagation network. Correspondingly, we define six categories of disinformation topics, each closely aligned with one specific community. Thus, both the number of communities and disinformation categories are uniformly denoted by $J$, while an individual community or disinformation type is indexed by $j$.

\section{Methodology}
\label{sec:B}
\subsection{Real-World Data}
 \label{sec:B.1}

\textbf{Data collection sources}: We select six representative vertical communities from the `Communities' module on the X/Twitter platform as data sources: Entertainment, Technology, Sports, Business, Politics, and Education. For each community, we systematically collect:

(1) Users’ basic profile information: user name, self-introduction of the user, number of posts by the user, number of followers, number of follows, creation time, whether verified or not.  

(2) A sample of 200 content posts, including original posts, retweets, and quotes.  

\textbf{Data filtering strategies}: Given that some users exhibit a high posting frequency—reaching the 200-post threshold within a short period—we apply a time window from February 7 to February 10, 2025 (a total of three days) to filter the data. This approach allows for the analysis of time-series characteristics such as user sharing frequency and activation times while ensuring consistency. Furthermore, to enhance simulation efficiency, we randomly sample approximately 100 users from the collected data of each community to participate in the subsequent simulations. As shown in Table \ref{table:1}, we provide detailed statistics of each community after applying the time-based filtering and user sampling, including:

(1) The number of users in each community.  

(2) The number of valid posts, retweets and quotes within the specified time frame.  

This data processing strategy ensures the reliability and comparability of our subsequent analysis.

\textbf{Potential limitations}: The collected dataset size is relatively small compared to real-world social networks, and our analysis covers only six communities, whereas the actual number of online communities is much larger. In addition, we filter only three days of user activity and limited the dataset to 200 text entries per user, which allows us to capture only short-term interests rather than long-term user preferences.

\subsection{Interest Community Prompt}
\label{sec:B.2}
Although Appendix \ref{sec:B.1} describes how users are collected from the "Communities" module on the X platform, this doesn't imply that the collected users are solely interested in the community from which they are sampled. Gathering users from the six major communities is to ensure the presence of active users within those domains, but it doesn't restrict their interests to just one community. Therefore, it's necessary for us to further evaluate the diversity of their interest communities. This step ensures that the subsequently constructed propagation network topology has multiple communities, and that a subset of users with cross-community interests can serve as a path for information transfer between these communities.

We design a prompt to evaluate users' interest communities based on their historical posts, retweets, quotes, and basic profile information. Below is the detailed prompt.

\begin{tcolorbox}[
  title= Interest Community Intensity Prompt,
  fonttitle=\large,
  center title,
  breakable]

  Please evaluate the user's interest intensity in different communities based on their [User Data], including historical posts, retweets, quotes, and basic profile information. The interest communities include \textcolor{blue}{Entertainment, Technology, Sports, Business, Politics, and Education}. Based on the provided user data, assign a score (\textcolor{blue}{1-10}) for each domain and briefly explain the reasoning behind the rating. If certain domains lack sufficient data support, please mark them as "\textcolor{blue}{Insufficient Data}" and explain the reason. Your response must strictly follow the [Output Format].

  \vspace{4pt}

  [\textcolor{blue}{User Data:}] 

  \textbf{\textcolor{black}{Personal Description:}} $\{personal\_description\}$  

  \textbf{\textcolor{black}{Historical Posts:}}$\{historical\_posts\}$

\textbf{\textcolor{black}{Historical Retweets:}}$\{historical\_retweets\}$

  \textbf{\textcolor{black}{Historical Quotes:}}$\{historical\_quotes\}$
  \vspace{4pt}

  [\textcolor{blue}{Evaluation Guidelines:}]

  1. Assign a rating of 1-10 points to each community

  2. Provide a brief rationale for each rating

  3. If there is insufficient data in a community, mark it as "Insufficient Data" and explain why.

  [\textcolor{blue}{Output Requirements:}]

  1. Strictly valid JSON format only

  2. No additional text outside the JSON structure

  3. All 6 communities must be included

  4. Use double quotes for all strings

  5. Escape special characters properly

  6. Response must strictly follow the [Output Format]

  [\textcolor{blue}{Output Format:}]  

  Required JSON Output Format:

  \{"Interest Community Scores": [

      \quad\{"Community": "Entertainment",

      \quad  "Score": "Score 1",

      \quad  "Reasoning": "Reasoning 1"\},

      \quad\{"Community": "Technology",

      \quad  "Score": "Score 2",

      \quad  "Reasoning": "Reasoning 2"\},

      \quad...

      \quad\{"Community": "Education",

      \quad  "Score": "Score 6",

      \quad  "Reasoning": "Reasoning 6"\}

    ]\}

\end{tcolorbox}

\begin{table*}[ht]
  \centering
  \caption{Parameter Definitions, Values, and Sources. In the “Value” column, values before the "/" represent the parameter range, and values after the "/" indicate the selected setting.} 
  \renewcommand\arraystretch{1.1}
  \setlength{\tabcolsep}{0.5mm}
  \adjustbox{width=0.99\textwidth}{%
  \begin{tabular}{c|c|c|c|c}
    \toprule[1pt]
    \textbf{Module} & \textbf{Parameter} & \textbf{Value}          & \textbf{Definition}                                                                        & \textbf{Source}            \\ \hline
    \multirow{8}{*}{Attributes} & $\mathcal{IC}_{uj}$       & {[}1, 10{]}     & The degree of interest of user $u$ in community $j$                               & LLM       \\
                              & $\mathcal{TT}_{uj}$       & {[}0, 1{]}      & The probability that user $u$ perceives information $j$ as disinformation.   & LLM       \\
                              & $\mathcal{DT}_{uj}$       & {[}0, 1{]}      & The tendency of user $u$ to disseminate information $j$      & Equation \ref{equ:1}      \\
                              & $\mathcal{SI}_{uj}$       & {[}0, 1{]}      & The social influence of user $u$ in community $j$      & Real data calculation     \\
                              & $\mathcal{AT}_{ut}$    & \{0.1, ...\}       & The probability that user $u$ is activated at time step $t$   & Real data statistics  \\
                              & $\theta$        & (0, 1)/0.5          & The balance parameter of interest community and truncated power-law score & Predefined      \\
                              & $\alpha$        & (1, $+\infty$) & The exponential part of the truncated power-law distribution                              & Real data fitting \\
                              & $\lambda$       & (0, $+\infty$) & The truncation intensity of the truncated power-law distribution in Eq. \ref{equ:1}        & Real data fitting \\
                              & $T$            & {[}1, 72{]}    & The length of the time step in the simulation                                     & Predefined      \\ \hline
    \multirow{1}{*}{Disinfo}  & $\mathcal{DP}_{j}$        & {[}0, 1{]}      & The plausibility score of the disinformation $j$                                  & LLM       \\ \hline
    \multirow{2}{*}{Bot List}  & $\mathcal{MB}_{j}$        & 0.15 * $N$        & The number of malicious bots in the dissemination network                                 & Predefined      \\
                                  & $\mathcal{LR}_{j}$        & 0.05 * $N$       & The number of legitimate bots in the dissemination network                                & Predefined      \\ 
                              & $\mathcal{MF}_j$  & [1, 18]   & The number of times the malicious bot is activated within the total simulated time step & Predefined \\
                              & $\mathcal{LF}_j$    & [1, 12] & The number of times the legitimate bot is activated within the total simulated time step  & Predefined \\ \hline
    \multirow{4}{*}{Network}     & $N$       & 689     & The number of dissemination network nodes    & Real data \\ 
                                & $m_0$       & 5       & The initial number of fully connected nodes in the BAM process   & Predefined \\
                                & $m$         & [1, 4]       & The number of edges introduced by newly added nodes in the BAM process  & Predefined \\
                                & $\tau$      & 8       & The threshold for the allocation of communities that users are interested in  & Predefined \\ \hline
    \multirow{3}{*}{Correct} & $\mathcal{EI}$            & {[}12, 72{]}     & The time step range of early intervention                                               & Predefined      \\
                                          & $\mathcal{MI}$            & {[}36, 72{]}   & The time step range for mid-term intervention                                           & Predefined      \\
                                          & $\mathcal{LI}$            & {[}48, 72{]}    & The time step range of late intervention                                                & Predefined      \\ \hline
    \multirow{12}{*}{Quantify} & $\gamma$        & (0, 1)/0.5          & The balance parameter between enhancement and decay term in Eq. \ref{equ:2}           & Predefined      \\
                                          & $\beta$         & (0, 1)/0.5          & The rate of change of the enhancement term in Eq. \ref{equ:2}      & Predefined      \\
                                          & $\delta$        & (0, 1)/0.5          & The rate of change of the decay term in Eq. \ref{equ:2}   & Predefined      \\
                                          & $F'_{kj}$       & {[}0, 1{]}      & The persuasiveness of corrective info $j$ towards user $k$ in Eq. \ref{equ:2}                & LLM       \\
                                          & $F_{kj}$        & {[}0, 1{]}      & The persuasiveness of disinfo $j$ towards user $k$ in Eq. \ref{equ:2}                     & LLM       \\
                                          & $\mathcal{\hat{TT}}_{uj}$ & {[}0, 1{]}      & The updated trust threshold of user $u$'s repeated exposure in community $j$              & Equation \ref{equ:2}      \\
                                          & $\mathcal{DA}_{uj}$       & {[}0, 1{]}      & The probability that user $u$ successfully identifies disinfo $j$       & Equation \ref{equ:3}      \\ 
                                          & $\mathcal{SR}_t^j$        & {[}0, 1{]}      &  Proportion of users unexposed to disinfo $j$ at time $t$    & Simulation changing  \\
                                          & $\mathcal{ER}_t^j$        & {[}0, 1{]}      & Proportion of users having encountered disinfo $j$ at least once at time $t$  & Simulation changing  \\
                                          & $\mathcal{IR}_t^j$        & {[}0, 1{]}      & Proportion of spreaders believe disinfo $j$ at time $t$       & Simulation changing  \\
                                          & $\mathcal{UR}_t^j$        & {[}0, 1{]}      & Proportion of spreaders unbelieve disinfo $j$ at time $t$      & Simulation changing  \\ \hline
  \end{tabular}}
  \label{table:2}
  \end{table*}

\subsection{Trust Threshold Prompt}
\label{sec:B.3}
We assess users' trust thresholds when encountering disinformation across different communities based on the semantic features of their historical posts, reposts, and comments, as well as their publicly available basic attributes. Our prompt design fully considers multiple dimensions of user-generated content, including linguistic style and sentiment tendencies, source citations, interaction patterns, and fundamental user attributes, ensuring a comprehensive and precise evaluation. The following is a detailed Prompt case.

\begin{tcolorbox}[
  title= Trust Threshold Evaluation Prompt,
  fonttitle=\large,
  center title,
  breakable]

  Please evaluate the user's trust threshold when encountering disinformation across different communities — \textcolor{blue}{Entertainment, Technology, Sports, Business, Politics, and Education} — based on [User Data], including historical posts, reposts, quotes, and basic attributes. The trust threshold represents a user's expectation regarding the truthfulness of received information. Provide a trust threshold score (0-1 scale) for each community, where:

  - [\textcolor{blue}{0.8-1.0}] = Highly resilient to disinformation

  - [\textcolor{blue}{0.5-0.79}] = Moderately resilient to disinformation

  - [\textcolor{blue}{0.0-0.49}] = Potentially vulnerable to disinformation
  
  [\textcolor{blue}{User Data:}]

  \textbf{- Basic Attributes:}

    • Follower Count:$\{follower\_count\}$

    • Following Count: $\{following\_count\}$

    • Personal Description: $\{personal\_description\}$

  \textbf{- Content Data:}  

    • Historical Posts: $\{historical\_posts\}$

    • Historical Retweets: $\{historical\_retweets\}$

    • Historical Quotes: $\{historical\_quotes\}$

    [\textcolor{blue}{Evaluation Guidelines:}]

    \textbf{1. Linguistic Style and Sentiment Tendencies: }

      - Critical thinking markers (e.g., "requires verification", "source needed")

      - Sentiment polarity distribution (skepticism vs. credulity markers)

      - Hedging language frequency
  
    \textbf{2. Education Level:}

      - Technical terminology accuracy

      - Complex sentence ratio

      - Academic reference frequency 
  
    \textbf{3. Source Reliability: }

      - Authoritative source citation rate (gov/academia)

      - Low-credibility source retweets
  
    \textbf{4. Basic Attributes: }

      - Personal descriptions, follower-to-following ratio analysis

      - Disinformation report history

      [\textcolor{blue}{Output Requirements:}]

    1. Strictly valid JSON format only

    2. No additional text outside the JSON structure

    3. All 6 communities must be included

    4. Use double quotes for all strings

    5. Escape special characters properly

    6. Trust threshold (0-1 scale)

    7. Response must strictly follow the [Output Format]

\vspace{4pt}

[\textcolor{blue}{Output Format:}]

Required JSON Output Format:

\{"Trust Threshold Scores": [

  \quad\{"Community": "Entertainment",

  \quad  "Score": "Score 1",

  \quad  "Reasoning": "Reasoning 1"\},

  \quad\{"Community": "Technology",

  \quad  "Score": "Score 2",

  \quad  "Reasoning": "Reasoning 2"\},

  \quad...

  \quad\{"Community": "Education",

  \quad  "Score": "Score 6",

  \quad  "Reasoning": "Reasoning 6"\}

  ]\}

\end{tcolorbox}

\subsection{Truncated Power-Law Distribution Fitting}

\label{sec:B.4}
Based on real-world data, we fit truncated power-law distribution functions $P_{uj} = C\cdot (x_u)^{-\alpha}e^{-\lambda}$, where $x_u$ represents the total number of retweets and quotes by a user. We employ maximum likelihood estimation (MLE) to optimize the truncated power-law exponent $\alpha$ and the lower bound $x_{min}$, subsequently calculating the normalization constant $C$ as $C = \frac{\alpha - 1}{x_{min}^{1-\alpha}}$. The exponential term $e^{-\lambda}$ acts as a cutoff to prevent infinite values. Our analysis of the collected real user data yielded a fitted distribution $P_{uj}(x_u) = 1.006 \cdot x_u^{-1.146}e^{-0.006},\  x_u \geq 16$. Recognizing biases in the real-world data, the fitted truncated power-law distribution function achieves higher accuracy for $x_i \geq 16$. To mitigate the impact of estimation errors for $x_i < 16$,  we reduce the  $\theta$ parameter in Equation \ref{equ:1}.

















\subsection{Disinformation Plausibility Prompt}
\label{sec:B.6}

\begin{tcolorbox}[
  title= Disinfo Plausibility Evaluation,
  fonttitle=\large,
  center title,
  breakable]

  Excluding all background knowledge, rigorously evaluate the plausibility of the [Input] below by analyzing:  

  1. \textbf{Emotional Expression} (e.g., exaggerated language, manipulative appeals)

  2. \textbf{Propaganda Strategies} (e.g., cherry-picking, strawman arguments)

  3. \textbf{Information Framing} (e.g., bias via omission, misleading context) 

  4. \textbf{Logical Consistency}: (e.g., contradictions, unsupported claims, or logical fallacies)

  [\textcolor{blue}{Input}]: $\{DisinformationText\}$

  [\textcolor{blue}{Output Requirements:}]

    1. Strictly valid JSON format optionally

    2. No additional text outside the JSON structure

    3. Use double quotes for all strings

    4. Response must strictly follow the [Output Format]
    
    [\textcolor{blue}{Output Format}]:

  Required JSON Output Format:

    \{
      \ "PlausibilityScore": [0-1],

      \quad"Reasoning": "brief explanation"\}

\end{tcolorbox}

\subsection{Disinformation Dissemination Network Construction}
\label{sec:B.7}

We employ the Stochastic Block Model (SBM) and the Barabási–Albert Model (BAM) to construct a disinformation dissemination network $\mathcal{G} = \{\mathcal{V}, \mathcal{E}\}$ that exhibits both scale-free properties and community structure. This approach ensures that the network follows to the preferential attachment mechanism while also reflecting the characteristic of dense intra-community connections and sparse inter-community links. The construction process consists of the following steps:

(1)\textbf{Node Community Assignment} (SBM Component): Given a network with $N$ users and $J$ communities, each user $u$ is assigned to one or more communities based on their community interest scores $\mathcal{IC}_{uj}$, which are evaluated by LLMs, and a predefined threshold $\tau$. Specifically:

\begin{itemize}
  \item 
If $\mathcal{IC}_{uj} \ge \tau$, user $u$ is assigned to community $j$, and thus belongs to the set of assigned communities $\{C_j\}$.

 \item 
Since a user's interest community scores $\mathcal{IC}_{uj}$ may exceed the threshold for multiple communities, they can be assigned to more than one community.
\end{itemize}

The threshold $\tau$ is designed to reflect the structural characteristic of strong intra-community connectivity and limited inter-community links, ensuring a realistic simulation of information diffusion patterns.

(2)\textbf{Initialize Intra-Community Connections} (BAM Component): Within each community $j$, we first construct a fully connected subgraph $\mathcal{G}_j = \{\mathcal{V}_j, \mathcal{E}_j\}$ for each community $j$, consisting of an initial set of $m_0$ seed nodes. New nodes $u$ are then iteratively introduced into the community, forming connections according to: 

\begin{itemize}
  \item
  The preferential attachment rule of the BAM.

  \item
  Add new nodes $u$ to community $j$ based on the social influence $\mathcal{SI}_{ej}$ of the existing nodes $e \in \mathcal{V}_j$ in the propagation network, where $\mathcal{SI}_{ej}$ is defined as:

            \[
              \mathcal{SI}_{ej} = \frac{F_e}{\sum_{i \in \mathcal{V}_j} F_i}
              \]
  $F_e$ represents the follower count (or another influence metric) of node $e$.

  This process leads to the formation of a scale-free network in which high-degree nodes (i.e., influential users) act as hubs for information dissemination, while low-degree nodes resemble ordinary users' social interaction patterns.
  
\end{itemize}

(3) \textbf{Iterative Network Growth}:

The iterative addition of new nodes $u$ continues until the community $j$ reaches its designated size. The complete procedure for constructing the disinformation dissemination network is detailed in Algorithm \ref{Al:1}.

\begin{algorithm*}[t!] 
  \caption{Building Disinformation Dissemination Networks Based on SBM-BAM}
  \label{Al:1}
  \begin{algorithmic}[1]
  \STATE \textbf{Specify the Inputs and Outputs}
  
  \STATE \textbf{Inputs:} Total nodes $N$. Initial number of nodes in each community $m_0$. Number of communities $J$. User interest community scores $\mathcal{IC}_{uj}$. Predefined threshold for community assignment $\tau$. Number of edges each new node forms $m, (m \le m_0)$.
  \STATE \textbf{Outputs:} Constructed dissemination network $\mathcal{G} = (\mathcal{V},\mathcal{E})$.
  
  \STATE \textbf{(1) Node Community Assignment (SBM Component):}
  \FOR {$u \leftarrow 1$ to $N$}
      \STATE Load $u$'s interest community scores $\mathcal{IC}_{uj}$ for all communities $j \in \{1, \ldots, J\}$
      \STATE Initialize set $C_j = \emptyset$ for all communities $j \in \{1, \ldots, J\}$
      \FOR {$j \leftarrow 1$ to $K$}
        \IF {$\mathcal{IC}_{uj} \ge \tau$}
            \STATE Add $u$ to node set of community $C_j$
        \ENDIF
      \ENDFOR
  \ENDFOR
  \STATE \textbf{(2) Initialize Intra-Community Connections (BAM Component):}
  \FOR {$j \leftarrow 1$ to $J$}
      \STATE Initialize community $j$'s graph $\mathcal{G}_j = \{\mathcal{V}_{j}, \mathcal{E}_j\}$, where $\mathcal{V}_{j}$ are the first $m_0$ users assigned to community $j$
      \STATE Create a complete graph on $\mathcal{V}_{j}$ and set $\mathcal{E}_j$ to these edges.
  \ENDFOR
  
  \STATE \textbf{(3) Iterative Network Growth (BAM Component):}
  \FOR {$j \leftarrow 1$ to $J$} 
          \FOR {$u$ in $C_j$}
            \FOR {existing node $e$ in $\mathcal{V}_{j}$}
              \STATE Compute social influence score: 
              \[
              \mathcal{SI}_e = \frac{F_e}{\sum_{i \in \mathcal{V}_{j}} F_i}
              \]
              \noindent where $F_e$ is the follower count of $e$.
            \ENDFOR
            \STATE Select $m$ nodes from $\mathcal{V}_j$ based on $\mathcal{SI}_e$ to generate $\mathcal{V}_{selected}$
            \FOR {each selected node $n$ in $\mathcal{V}_{selected}$}
                \STATE Add edge $(n,u)$ to $\mathcal{E}_j$.
            \ENDFOR
            \STATE Add node $u$ to $\mathcal{V}_j$.
          \ENDFOR
    \ENDFOR
  
  \RETURN The network $\mathcal{G} = (\mathcal{V},\mathcal{E})$.
  \end{algorithmic}
\end{algorithm*}

\subsection{Corrective Strategy Example}

\label{sec:B.8}

We use an example of disinformation in the political community to illustrate how two correction strategies can be employed to curb the spread of disinformation and guide the public toward accurate information.

\textbf{Disinformation Case}: "Trump said the House select committee that investigated the Jan. 6, 2021, Capitol attack deleted and destroyed all of the information that they collected over two years."

\textbf{Fact-based Corrective Strategy}: Correcting disinformation by conveying officially released statements.

\begin{tcolorbox}[
  title= Fact-based Corrective Strategy,
  fonttitle=\large,
  center title,
  breakable]

  \textcolor{blue}{Case 1}" The House Select Committee that investigated the Jan. 6, 2021, attack on the U.S. Capitol publicly released a final 845-page report and more than 100 transcripts of testimony, along with memos, depositions and documents. Much of the committee’s work remains available online. 
  
  \vspace{10pt}

  \textcolor{blue}{Case 2}: House Republicans in March 2024 released a report accusing the committee of suppressing and deleting some pieces of evidence, but they have not gone as far as Trump in claiming that "all" evidence was destroyed.

\end{tcolorbox}

\textbf{Narrative-based Corrective Strategy}: Correcting disinformation by sharing personal experiences of relevant individuals.

\begin{tcolorbox}[
  title= Narrative-based Corrective Strategy,
  fonttitle=\large,
  center title,
  breakable]

  \textcolor{blue}{Case 1}: The House committee members, consisting of seven Democrats and two Republicans, stated that certain videos and sensitive materials were excluded from the official record to protect witnesses. All other materials were properly preserved, including over 100 testimony transcripts, written depositions, and documents that remain publicly available online. 

  \vspace{10pt}

  \textcolor{blue}{Case 2}: "As a member of the committee, we were under tremendous pressure in those days," a Democratic MP recalled, with a hint of exhaustion. "In order to protect those who bravely stood up to testify, we had to make difficult decisions, and rule out extremely sensitive videos and materials. I always remember a young witness who trembled and burst into tears when she described the experience. Imagine what secondary harm would have caused her if those images were revealed? We must protect her, which is our most basic responsibility. As for other records, I can assure you that, including those over a hundred testimonies, and those publicly written records and documents, each kept intact, and anyone can check and learn the truth."
  


\end{tcolorbox}

\subsection{Intervention Timing Setting}
\label{sec:B.9}
\textbf{Early Intervention}: Early intervention initiates corrective actions shortly after the onset of disinformation spread, aiming to limit its initial propagation and exposure before it gains significant traction.

\textbf{Mid-Stage Intervention}: Mid-stage intervention implements corrective actions after some disinformation spread. This allows observation of initial diffusion patterns and targeted correction of affected individuals and influential spreaders to mitigate further amplification.

\textbf{Late Intervention}: Late intervention deploys corrective measures when disinformation has already been circulating within the network for an extended period. Although the disinformation may have established a stronger foothold by this stage, the late intervention still aims to: reduce its persistence, correct the beliefs, and mitigate long-term negative impacts.

\vspace{-5pt}
\subsection{Persuasiveness Evaluation Prompt}
\label{sec:B.10}

\begin{tcolorbox}[
  title= Persuasiveness Evaluation Prompt,
  fonttitle=\large,
  center title,
  breakable]
  You are a social network user. The summary of the content you have posted or forwarded and quoted in the past three days is as follows: 

  [\textcolor{blue}{history Info}]: $\{history\_info\}$

  Evaluate the persuasiveness of the given [Input] information for you. The evaluation must result in a normalized score between [0,1] with clear justification.

  [\textcolor{blue}{Input}]: $\{text\_information\}$

  Your assessment should synthesize the following dimensions and give normalized scores between [\textcolor{blue}{0–1}] and brief reasons:

  1. Relevance: how well it fits with your interests, career, or topics of concern

  2. Bias: Whether the information is highly consistent with your views or is deliberately biased

  3. Novelty $\&$ Informativeness: Whether the new ideas or data are provided compared to existing knowledge

  [\textcolor{blue}{Output Requirements:}]

  1. Assign a score of [0,1] points

  2. Include both quantitative score and brief reasoning

  3. Strictly valid JSON format only

  4. No additional text outside the JSON structure

  5. Use double quotes for all strings

  6. Escape special characters properly

  7. Response must strictly follow the [Output Format]

  [\textcolor{blue}{Output Format:}]

  Required JSON Output Format:

  \{
    \ "Score": [0,1],

    \quad "Reasoning": "brief\_explanation"
  \}

\end{tcolorbox}

\subsection{Simulation and Evaluation Algorithm}
\label{sec:B.11}

We present the simulation workflow for disinformation in the MADD framework in Algorithm \ref{Al:2}. The susceptible ratio (SR), exposed ratio (ER), infection spreader ratio (IR), and uninfection spreader ratio (UR) are obtained by calculating the proportion of susceptible users (SU), exposed users (EU), infection spreaders (IS), and uninfection spreaders (US) as determined by Algorithm 2, out of the total number of users.

\subsection{Interaction and Decision-Making of Agents}
\label{sec:B.12}

\textbf{Interaction and Decision-Making Strategy of Malicious and Legitimate Bots}:
 We assign each bot agent the same five categories of attribute information as human users, including a predefined interest community. In terms of propagation behavior, the bots exhibit the following characteristics: their trust threshold and dissemination tendency are both set to 1, ensuring that they always trust and actively propagate either disinformation (for malicious bots) or corrective information (for legitimate bots). Their social influence is randomly sampled to match the influence levels of certain human users, and their active time is determined based on the predefined time range specified in Table \ref{table:2}.
 The bots’ action policy is straightforward: once their activation time is reached, malicious bots will proactively disseminate misinformation, while legitimate bots will distribute corrective content.

\textbf{Interaction and Decision-Making Strategy of Human Users}:
 During their activation time, human users receive and read information propagated by their neighboring nodes. Their decision-making process is governed: (1) their dissemination tendency determines whether they are willing to share the received information; and (2) their trust threshold is used to evaluate whether they trust the content. On the premise of confirming whether the information is trusted, users will choose one of two dissemination methods—either retweeting or quoting—to share the content. This strategy is designed to realistically simulate human interaction behaviors and cognitive judgment processes in the context of information diffusion.

\section{Experimental Results}

\label{sec:C}
\subsection{MADD and Real-World Consistency Evaluation}
\label{sec:C.1}

\begin{figure}[]
  \centering
  \includegraphics[width=0.49\textwidth]{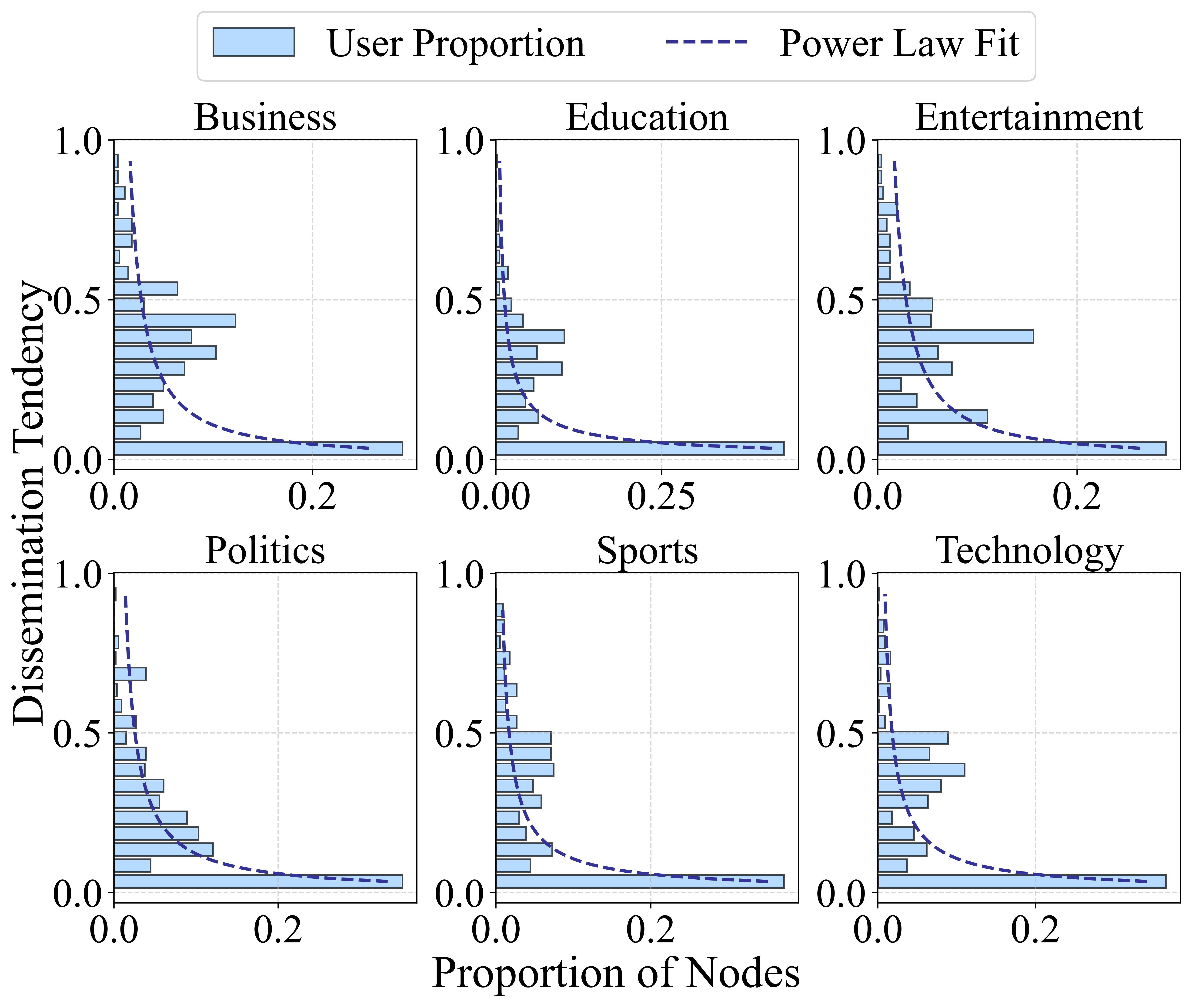}
  \caption{User dissemination tendency exhibits a heavy-tailed pattern.}
  \label{fig:4}
\end{figure}

Figure \ref{fig:4} demonstrates that the probability distribution of users' dissemination tendencies exhibits a distinct heavy-tailed characteristic: while the majority of users show low propagation activity, a small subset displays exceptionally high dissemination propensity. Although the relatively limited sample size may introduce some outliers, the overall distribution pattern remains consistent with previously documented user dissemination behaviors in the literature \cite{goel2016structural}.

\begin{figure}[htb]
  \centering
  \includegraphics[width=0.45\textwidth]{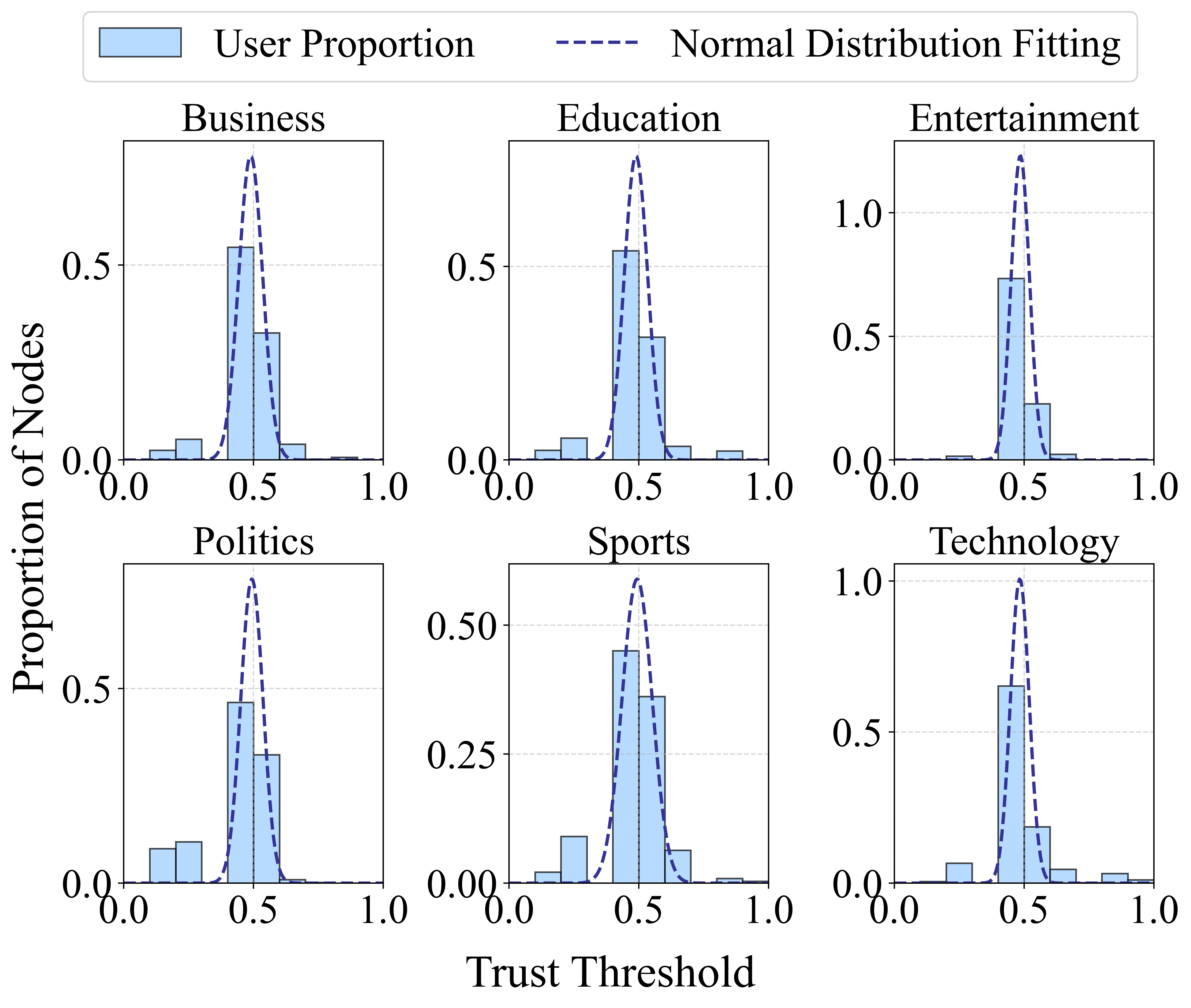}
  \caption{Trust threshold follow normal distribution fitting.}
  \label{fig:5}
\end{figure}

Figure \ref{fig:5} presents the distribution of users' trust thresholds toward disinformation. For all topic categories, the trust threshold distribution follows an approximately normal pattern, indicating that most users adopt a neutral stance of "neither fully believing nor completely rejecting" towards disinformation, with trust threshold scores predominantly clustered around $0.5$. This finding is consistent with established research outcomes \cite{ecker2022psychological}.

\subsection{Group-Level: Effectiveness Evaluation of Correction Strategies}
\label{sec:C.2}

\begin{figure}[htb]
  \centering
  \includegraphics[width=0.49\textwidth]{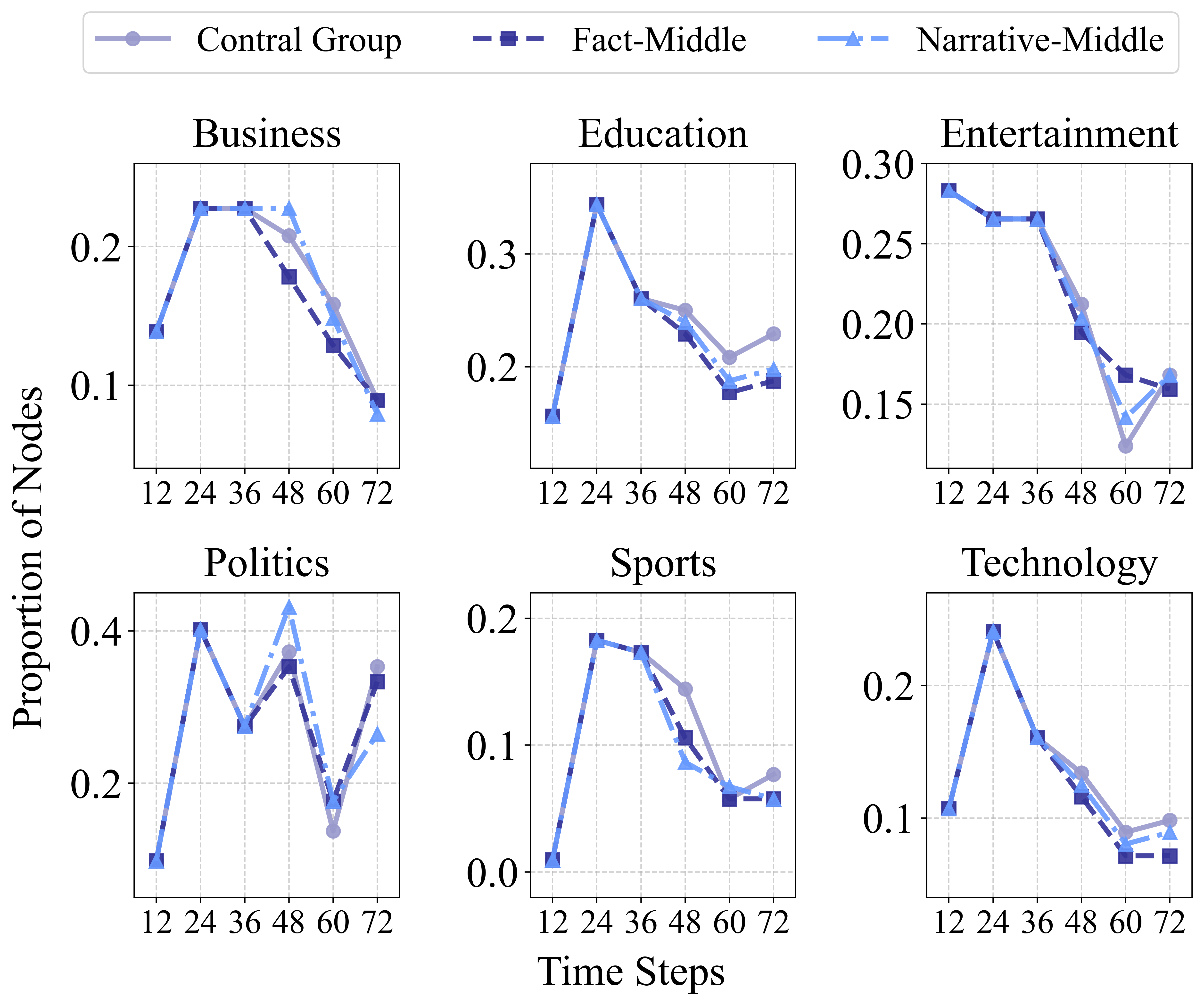}
  \caption{Comparison of mid intervention effects of two correction strategies.}
  \label{fig:10}
\end{figure}

\begin{figure}[htb]
  \centering
  \includegraphics[width=0.49\textwidth]{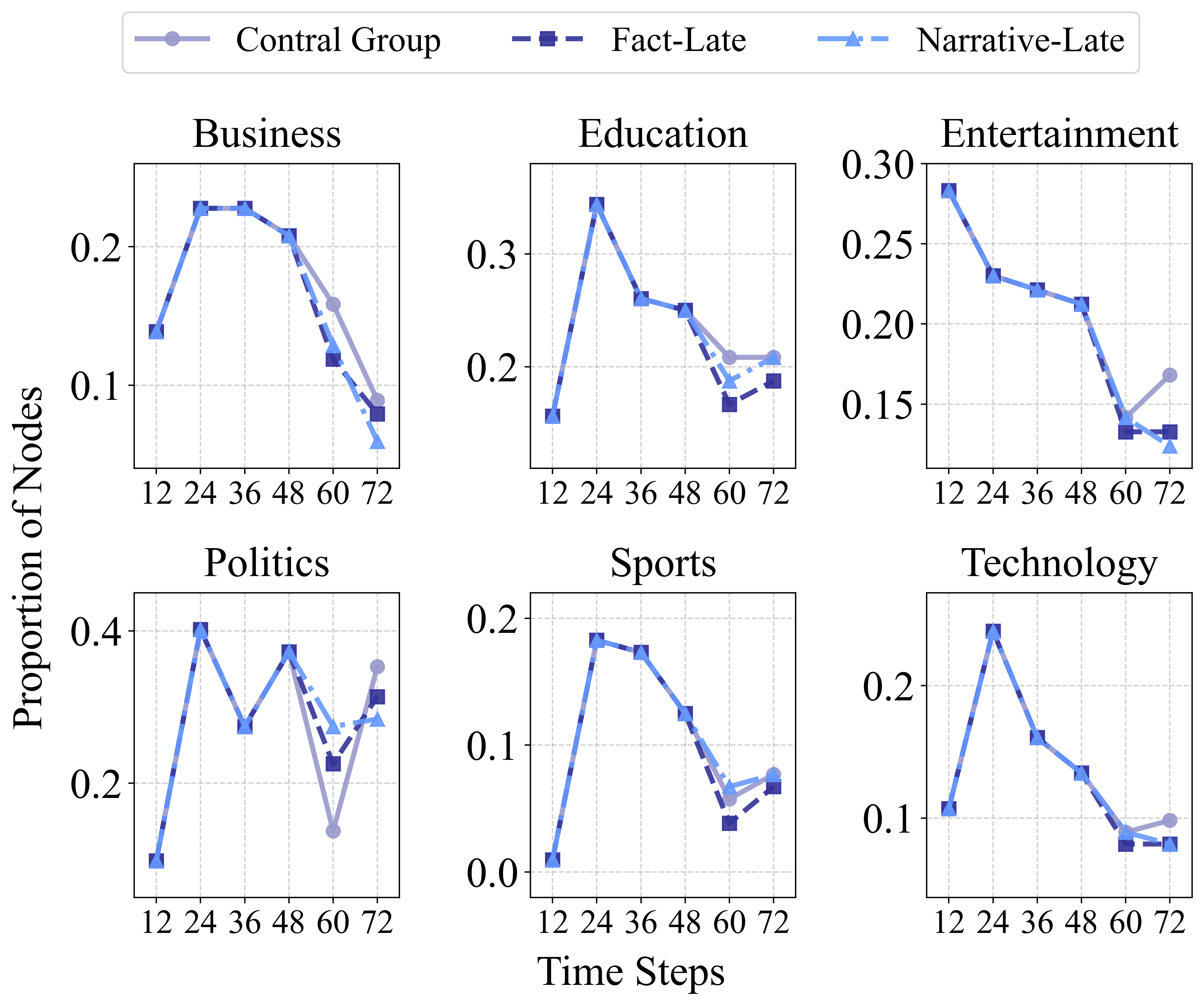}
  \caption{Comparison of late intervention effects of two correction strategies.}
  \label{fig:11}
\end{figure}

Figures \ref{fig:10} and \ref{fig:11} respectively illustrate the comparative intervention effects of two correction strategies (fact-based and narrative-based correction) implemented during the mid (T=[36-72]) and late (T=[48-72]) stages of disinformation spread. The results indicate that the intervention effects of both correction strategies are not significant for most communities. Notably, in some communities, such as "Politics" and "Business," we observe negative intervention effects, manifested as an increase in the spread of related disinformation after correction. This phenomenon may be because prolonged exposure to disinformation leads users to form stable incorrect cognitive frameworks, and the introduction of corrective information is interpreted by some users as a "threat to their original viewpoints," thereby triggering a psychological reactance effect \cite{diaz2023disinformation} and forming an echo chamber effect \cite{garimella2018political}.

\subsection{Individual-Level: Dynamic Changes in User Trust Thresholds}
\label{sec:C.3}

\begin{figure}[htb]
  \centering
  \includegraphics[width=0.49\textwidth]{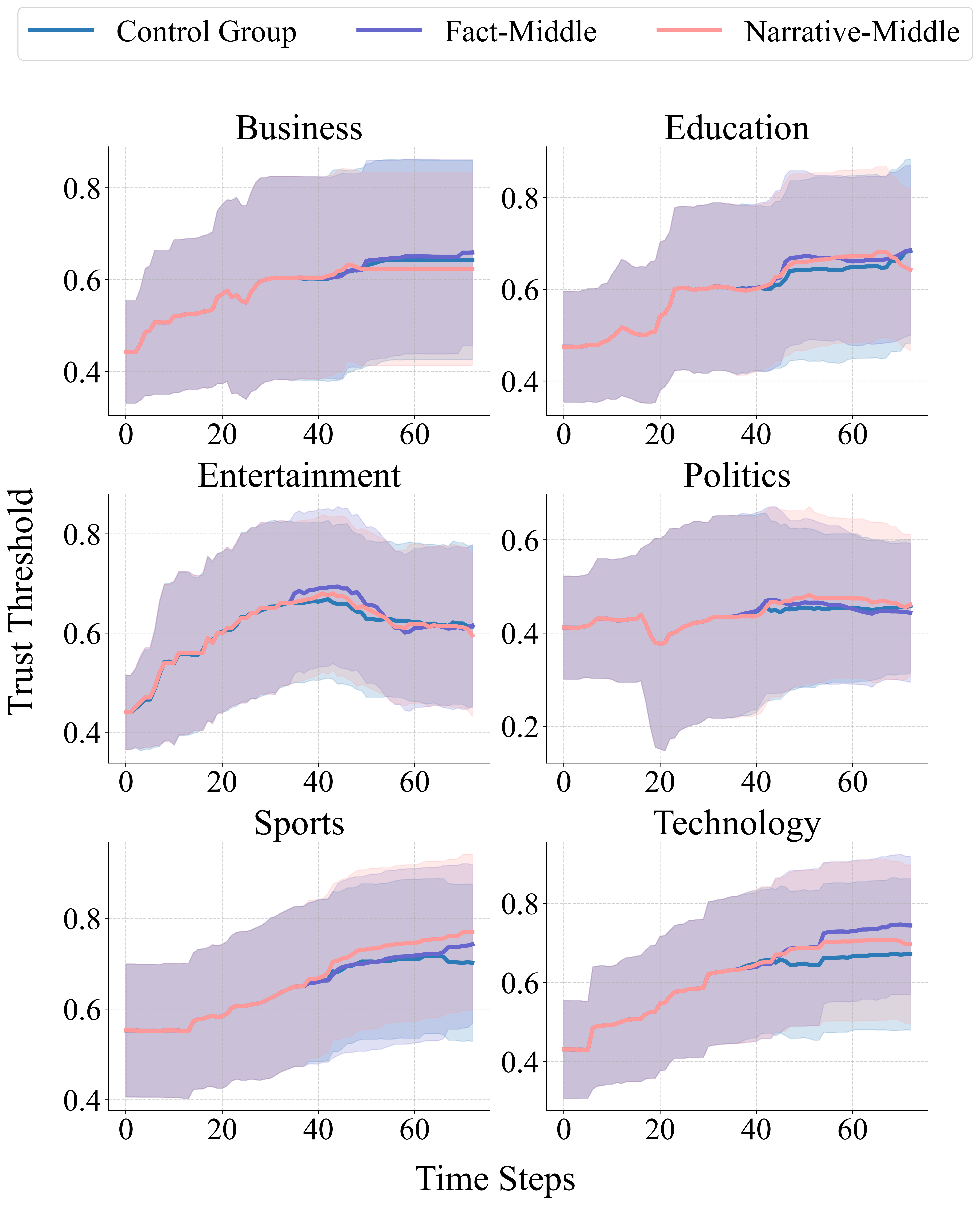}
  \caption{Effects of Correction Strategies on User Trust Thresholds During Mid-Stage Intervention}
  \label{fig:13}
\end{figure}

\begin{figure}[htb]
  \centering
  \includegraphics[width=0.49\textwidth]{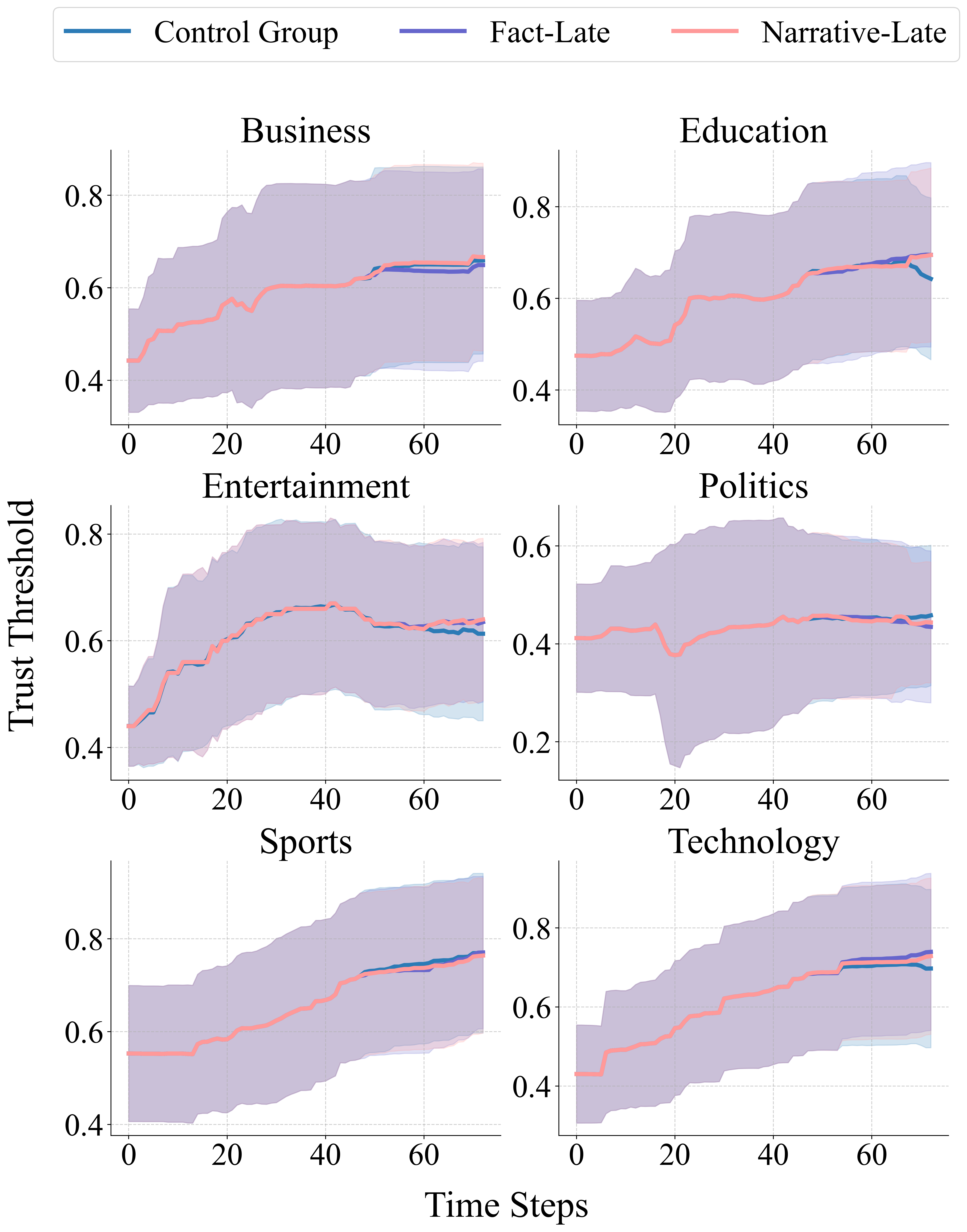}
  \caption{Effects of Correction Strategies on User Trust Thresholds During Late-Stage Intervention}
  \label{fig:14}
\end{figure}

Figures \ref{fig:13} and \ref{fig:14} illustrate the changes in user trust thresholds when correction strategies are implemented in the mid and late stages.  We observe that their effect on increasing user trust thresholds for evaluations is extremely limited. Furthermore, in the "Entertainment," "Business," and "Politics" communities during the mid-term intervention phase, and in the "Sports" community during the late-term intervention phase, the control group (without correction strategies) actually shows higher user trust thresholds than the experimental group (with correction strategies). This phenomenon likely stems from two main reasons: firstly, the intervention timing is delayed, and secondly, echo chamber effects may have already formed within some user groups. When disinformation has already spread widely across the network and deeply influenced users' cognitive belief systems, correction strategies introduced at this point often struggle to effectively reverse established negative perceptions. Therefore, these findings strongly confirm the critical importance of intervening in the early stages of disinformation spread and implementing effective correction strategies.

\subsection{Case Study: Change of the Trust Threshold of the Case User}

\begin{figure*}[htb]
  \centering
  \includegraphics[width=0.95\textwidth]{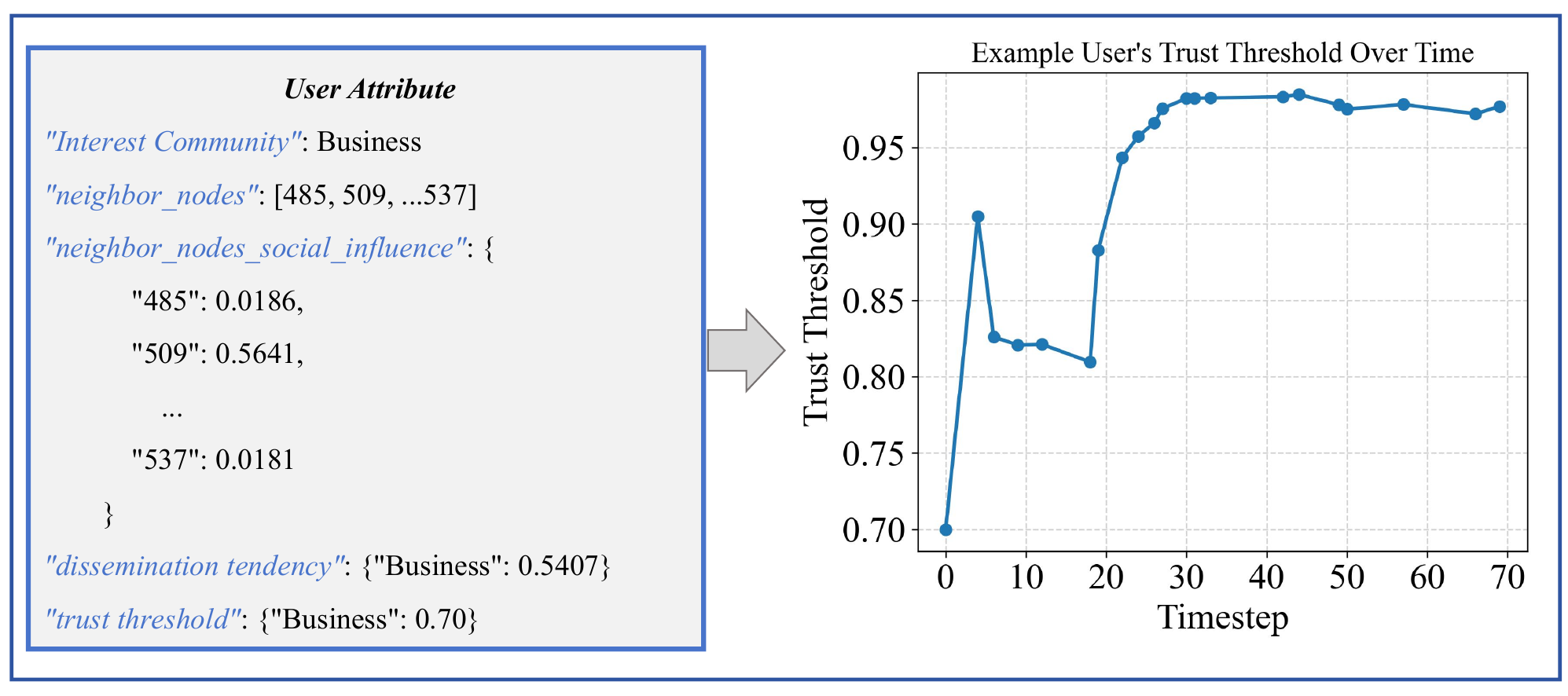}
  \caption{The changing trend of the trust threshold of the case users}
  \label{fig:16}
\end{figure*}

In Figure \ref{fig:16}, we show an example of a user's Trust Threshold over time under a fact-based early intervention strategy. Overall, the trust threshold undergoes a dynamic change: first rising sharply, then declining, and finally stabilizing.

The initial rapid increase is primarily because the user receives highly credible corrective information from high-influence nodes, which significantly enhances their ability to discern information. However, as disinformation gradually increases in the network, the user's trust threshold begins to decline slowly, reflecting some cognitive interference. As legitimate bots continue their correction efforts in the later stages, the proportion of true information in the user's environment increases, causing the trust threshold to rise again and eventually stabilize. This indicates that the user has developed a strong resistance to disinformation on this topic.

\subsection{Effect of Legitimate Bot Ratio on Correction}

Figure \ref{fig:17} illustrates the infection rate dynamics from time steps 12 to 72 under a fact-based early intervention strategy with varying proportions of legitimate bots. 

Overall, the infection rate consistently declines over time, but the speed and scale of this decline are directly tied to the strength of the intervention.

\begin{itemize}
    \item \textbf{No Intervention (0\% L-Bots):} Without any legitimate bots, the infection rate remains high (around 22.7\%) and declines slowly. This highlights that users' natural ability to correct disinformation is limited when there is no active intervention.

    \item \textbf{With Intervention (5\% to 30\% L-Bots):} As the proportion of legitimate bots increases from 5\% to 30\%, the infection rate drops much faster during the middle stages (time steps 24 to 48), proving the effectiveness of stronger corrective intervention.
\end{itemize}

Interestingly, the infection rates across all three intervention scenarios converge in the later stages (time steps 48 to 72). This is likely because the remaining infected users have low initial confidence, making them less susceptible to change. Furthermore, since bot intervention is concentrated in the early stages, the later dynamics are driven more by {users' own cognitive adjustments, leading to a similar rate of change across all groups.

\begin{figure}[htb]
  \centering
  \includegraphics[width=0.49\textwidth]{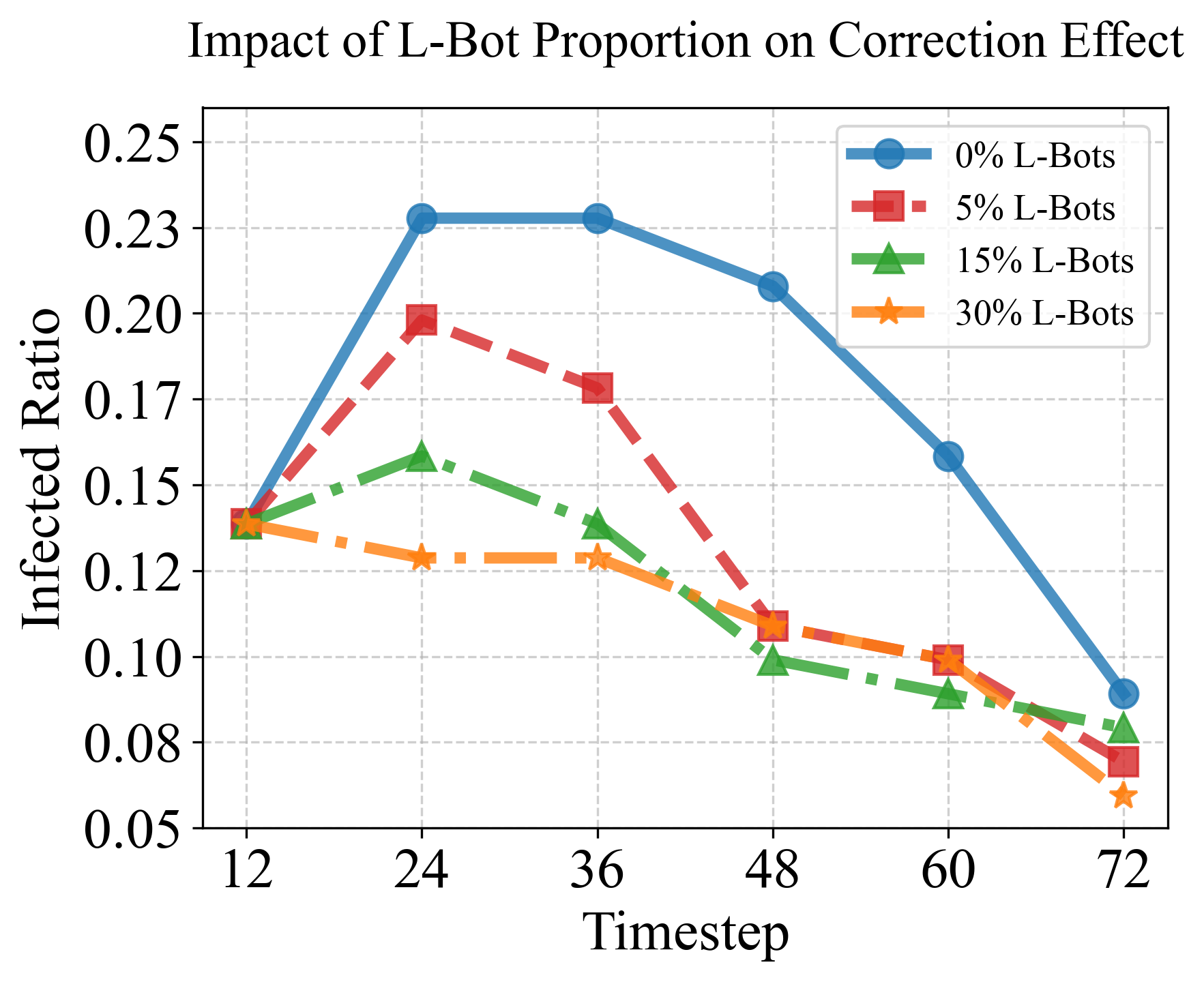}
  \caption{Effect of Legitimate Bot Ratio on Correction}
  \label{fig:17}
\end{figure}

\subsection{Long-Term Simulation}

To examine the long-term dynamics of user states and trust thresholds within the community, we conduct an extended behavioral simulation on the `Business' community, implementing an early fact-based intervention strategy. Specifically, we extend the simulation to 120 time steps to track changes in user states and trust thresholds, as illustrated in Figure \ref{fig:18}.

The results indicate that the long-term intervention significantly reduce the proportion of users influenced by disinformation, keeping it below 5\%. Meanwhile, the average trust threshold of users increased markedly from approximately 0.45 to nearly 0.72, suggesting a substantial enhancement in users’ ability to identify disinformation related to this topic. These findings validate that early fact-based interventions can effectively mitigate the influence of disinformation in prolonged information diffusion scenarios.

\begin{figure*}[h]
  \centering
  \subfigure[State Transition]{
    \begin{minipage}[t]{0.45\textwidth}
        \centering
        \hspace{-0.2cm}
        \includegraphics[width=0.98\textwidth]{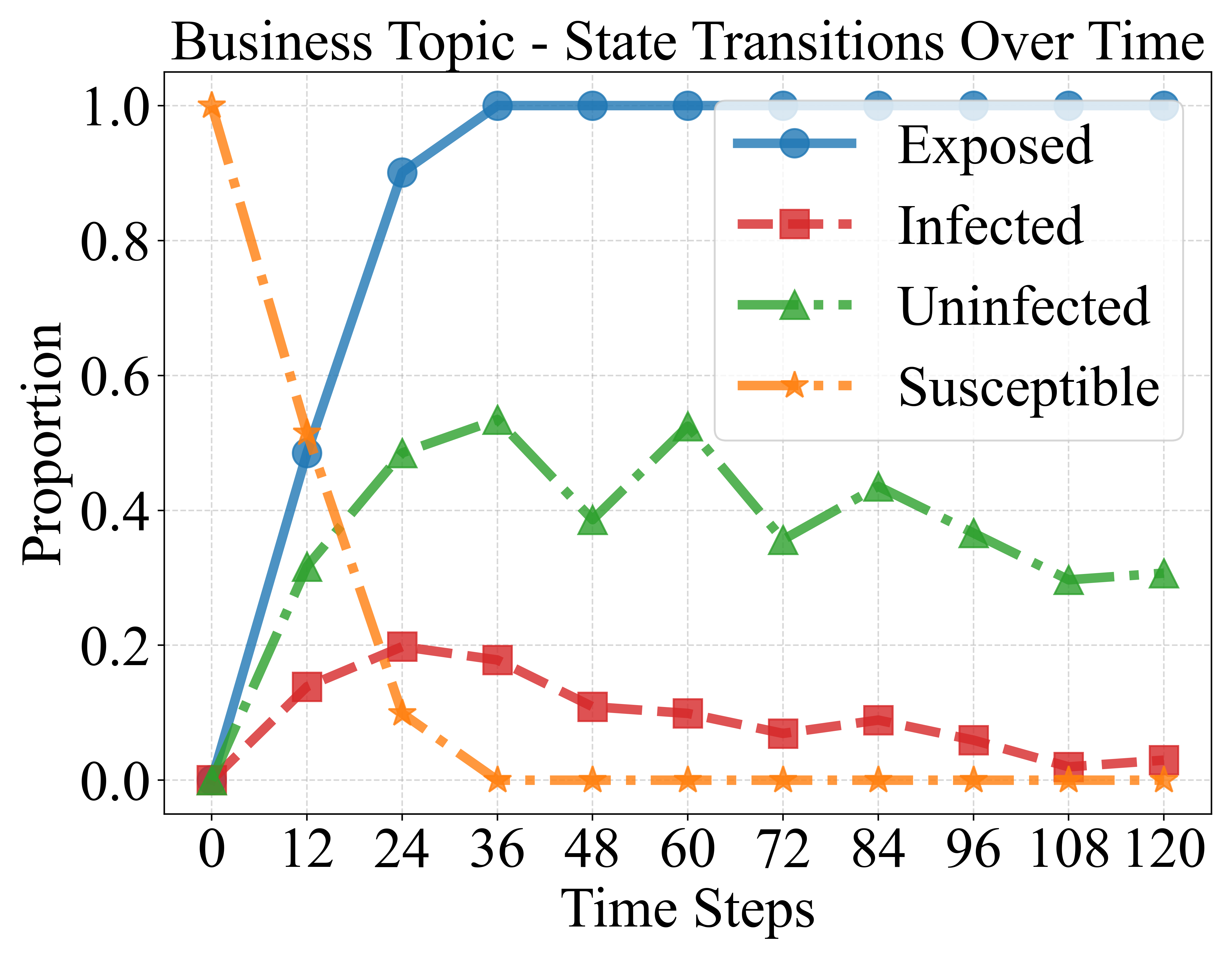}
    \end{minipage}
  }%
  \subfigure[Trust Thresholds.]{
    \begin{minipage}[t]{0.45\textwidth}
        \centering
        \hspace{-0.2cm}
        \includegraphics[width=0.98\textwidth]{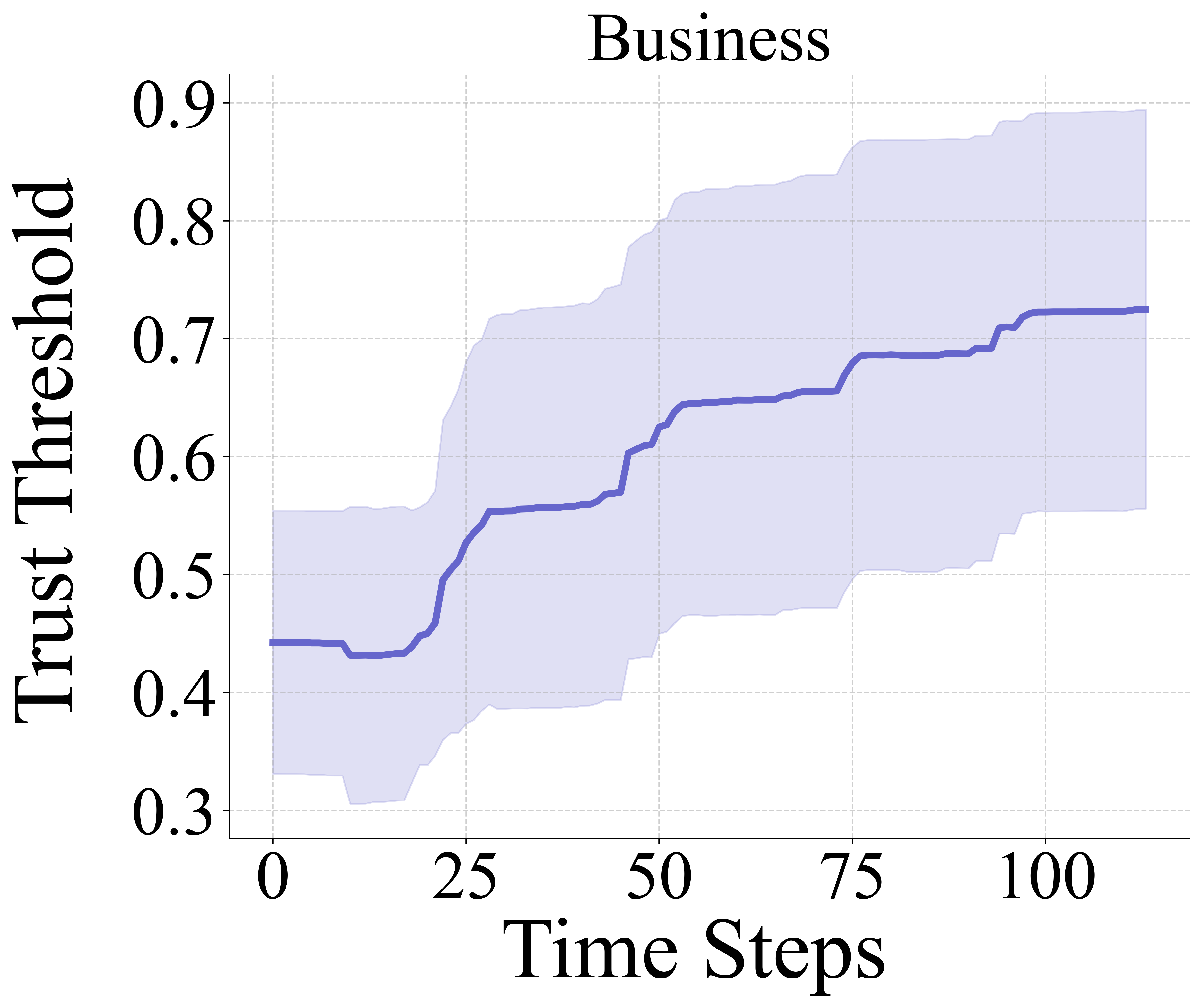}
    \end{minipage}
  }%
  \caption{Long-Term dynamics of User States and Trust Thresholds}
  \label{fig:18}
\end{figure*}

\subsection{Community Diffusion Analysis}


We analyze the global diffusion process across communities, as illustrated in Figure \ref{fig:19}. The heatmap on the left shows the number of users that exist independently within each community, as well as the number of overlapping users between different communities. The right panel presents the incoming time and total exposure time of users in other communities when different communities serve as the origin of disinformation dissemination. We observe that some communities are not fully exposed during the global simulation. This is primarily because these communities contain user nodes that are either less interested in the simulated topic or exhibit low activity levels. Such nodes tend not to propagate information, which in turn prevents their connected neighbors from becoming exposed.

\begin{figure*}[htb]
  \centering
  \includegraphics[width=0.95\textwidth]{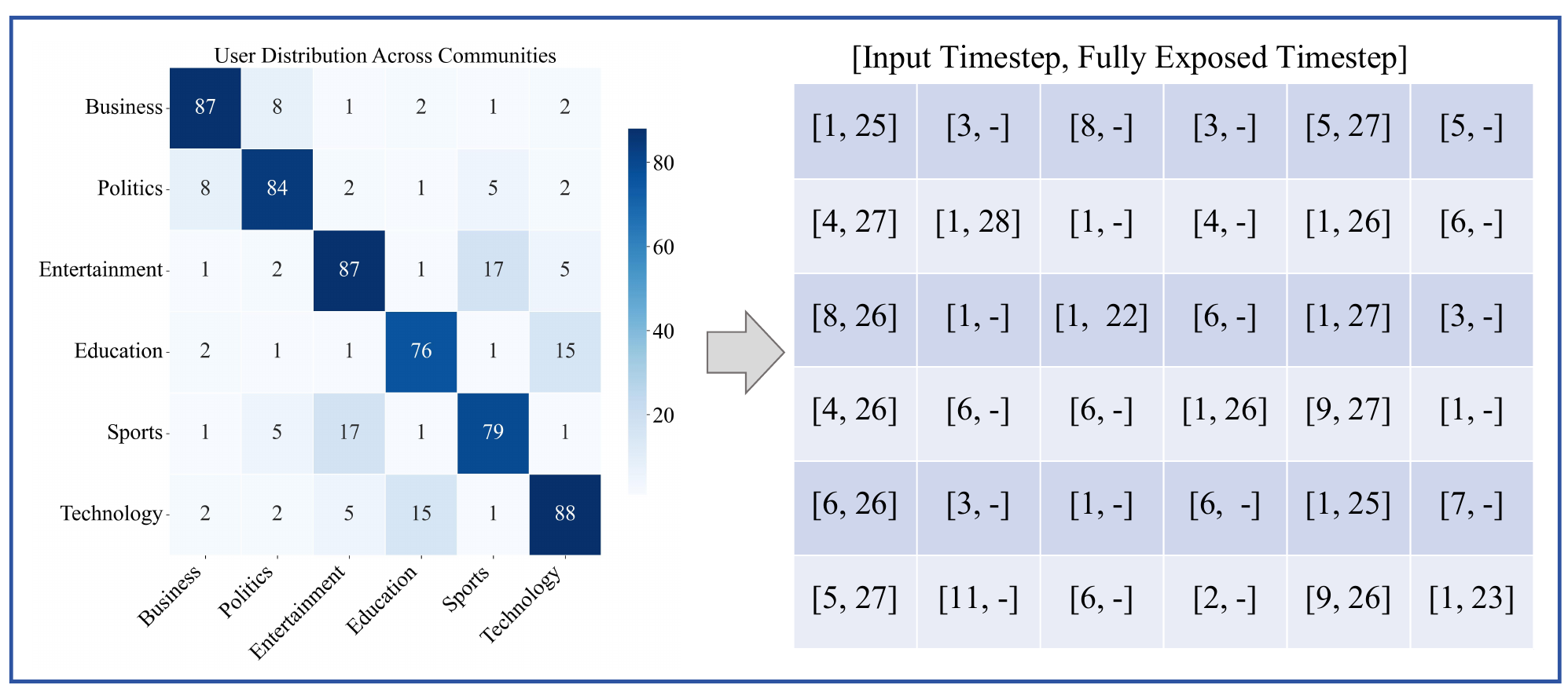}
  \caption{Heatmap of community distribution and Inter-community diffusion time steps. (a)Heatmap of community distribution: illustrates the number of users within each community and the overlaps between communities. (b) Inter-community diffusion time steps: indicate the number of time steps required for disinformation to spread from one community to another, including the initial exposure time and the time when all users are exposed.}
  \label{fig:19}
\end{figure*}

\begin{figure*}[htb]
  \centering
  \includegraphics[width=0.95\textwidth]{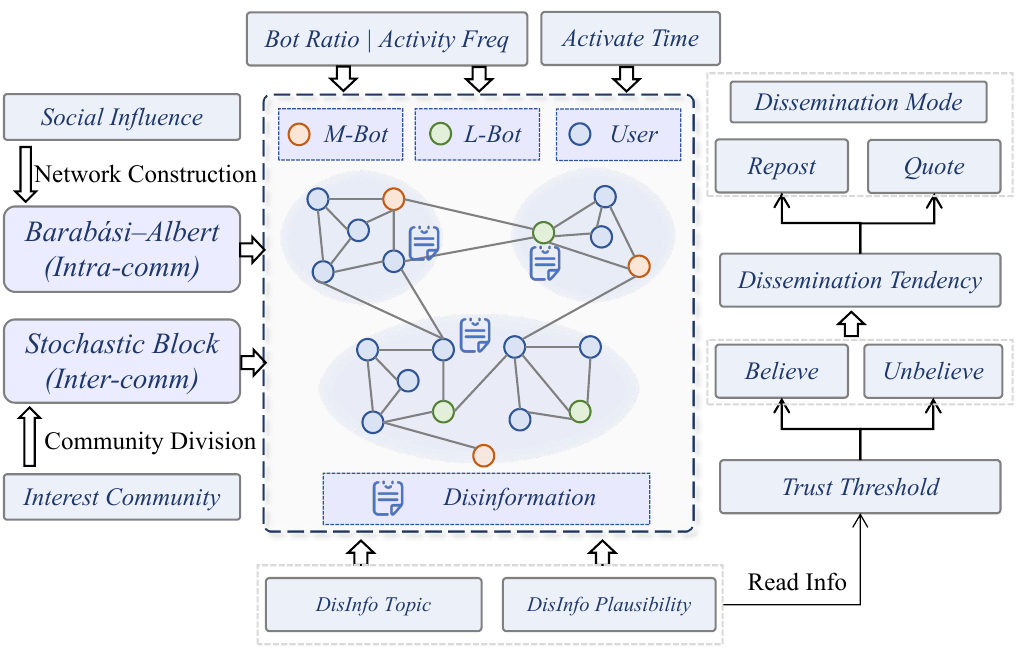}
  \caption{Supplementary expansion of Figure \ref{fig:2}. Attribute-driven network dynamics and disinformation dissemination in MADD.}
  \label{fig:20}
\end{figure*}

\subsection{System Resources Analysis}

We present in Table \ref{table:3} the key metrics for the control group (simulation involving only malicious bots), including the number of LLM calls, inference time, and token usage. Since the simulation relies on external LLM APIs, response speeds may vary, and inference time is not strictly proportional to the number of calls.

\begin{table*}[ht]
\centering
\caption{Statistics on the consumption of simulated resources for each community (LLM call volume, inference time, Token consumption)}
\renewcommand\arraystretch{1.2}
  \adjustbox{width=0.95\textwidth}{%
\begin{tabular}{lcccccc}
\toprule
\textbf{Metric} &
\textbf{Politics} &
\textbf{Business} &
\textbf{Education} &
\textbf{Entertainment} &
\textbf{Sports} &
\textbf{Technology} \\
\midrule
LLM Calls / Avg.\ per User
    & 374{,}212 / 366.9
    & 275{,}478 / 272.8
    & 259{,}344 / 2{,}702
    & 325{,}702 / 2{,}882
    & 256{,}483 / 2{,}466
    & 234{,}723 / 2{,}096 \\

Inference Time / Avg.\ per User
    & 19.02\,h / 671\,s
    & 17.53\,h / 625\,s
    & 16.82\,h / 631\,s
    & 14.73\,h / 469\,s
    & 12.59\,h / 435\,s
    & 12.27\,h / 394\,s \\

Token Usage / Avg.\ per User
    & 9{,}983{,}935 / 97{,}881
    & 8{,}773{,}767 / 86{,}869
    & 7{,}684{,}069 / 80{,}042
    & 9{,}893{,}845 / 87{,}556
    & 7{,}115{,}933 / 68{,}422
    & 7{,}503{,}570 / 66{,}996 \\
\bottomrule
\end{tabular}}
\label{table:3}
\end{table*}

\begin{algorithm*}[] 
  \centering
  \caption{Simulation and Evaluation Process}
  \label{Al:2}
  \begin{algorithmic}[1]
    \STATE \textbf{Inputs:} Time steps $\mathcal{T}$, Disinfo set $\mathcal{J}$, Corrective info set $\mathcal{C}$, Malicious Bots $\mathcal{MB}$, Legitimate Bots $\mathcal{LB}$, Regular Users $\mathcal{RU}$, Active users at time $t$, $\mathcal{AC}_t$, Initial received/shared info ($\mathcal{R}_u$, $\mathcal{S}_u$ for each user $u$).
    \STATE \textbf{Outputs:} Trust thresholds $\mathcal{TT}$, Susceptible users $\mathcal{SU}$, Exposed users $\mathcal{EU}$, Infection Spreaders $\mathcal{IS}$, Uninfection Spreaders $\mathcal{US}$.
    \STATE \textbf{Initialize:} $\mathcal{SU} \gets \mathcal{RU}$, $\mathcal{EU} \gets \emptyset$, $\mathcal{IS} \gets \emptyset$, $\mathcal{US} \gets \emptyset$
    \STATE \textbf{Start Simulation:}
    \FOR {disinfo $j \in \mathcal{J}$}
      \FOR {each time $t \in \mathcal{T}$}
        \FOR {each active user $u \in \mathcal{AC}_t$}
          \STATE \textbf{1. Load Share Info:}
          \IF {$u \in \mathcal{MB}$}
            \STATE Spread disinfo $\hat{info} = j$ to neighbors $\mathcal{N}_u$
          \ELSIF {$u \in \mathcal{LB}$}
            \STATE Spread corrective info $\hat{info} = c, c \in \mathcal{C}$ to neighbors $\mathcal{N}_u$
          \ELSIF {$(u \in \mathcal{RU}) \land (\mathcal{R}_u \neq \emptyset)$ }
            \STATE $info \gets$ Lastest received info ($\mathcal{R}_u$)
            \IF {random( ) < $\mathcal{DT}_{u,j}$}
              \STATE Action $\hat{info} \gets \mathcal{LLM}(info, history, Action List)$
            \ENDIF
            \STATE \textbf{2. Share Info:}
            \STATE Share $\hat{info}$ to neighbors $\mathcal{N}_u$, update $\mathcal{S}_u \gets \mathcal{S}_u \cup \{\hat{info}\}$
          \ENDIF
          \FOR {each neighbor $v \in \mathcal{N}_u \cap \mathcal{RU}$}
            \STATE \textbf{3. Receive Info:}
            \STATE Update $\mathcal{R}_v \gets \mathcal{R}_v \cup \hat{info}$
            \STATE Discern Ability $\mathcal{DA}_{v, t}^{\hat{info}} \gets \mathcal{LLM}(\text{Check Belief}(\mathcal{TT}_u, \mathcal{DP}_{\hat{info}}))$ based on Eq. \ref{equ:3}
            \STATE Update $\mathcal{TT}_{v,j}$ via:
            \STATE \quad $\mathcal{TT}_{v,j} \gets \text{Eq. } \ref{equ:2} \text{ Dynamic adjustment}$
            \STATE \quad Update state sets $\mathcal{SU}_{t,j}$, $\mathcal{EU}_{t,j}$, $\mathcal{IS}_{t,j}$, $\mathcal{US}_{t,j}$
          \ENDFOR
        \ENDFOR
      \ENDFOR
    \ENDFOR  
    \STATE \RETURN $\mathcal{SU}$, $\mathcal{EU}$, $\mathcal{IS}$, $\mathcal{US}$, $\mathcal{TT}$
  \end{algorithmic}
\end{algorithm*}

\end{document}